\documentclass[12pt,showkeys,pra,superscriptaddress]{revtex4-1}

\usepackage{amssymb}
\usepackage{amsfonts}
\usepackage{amsmath}

\usepackage{dcolumn}
\usepackage{bm}
\usepackage{epsfig}
\usepackage{epsf}
\usepackage[]{graphicx}
\usepackage[sort&compress]{natbib}
\usepackage{color}
\usepackage{xspace}

\renewcommand{\d}{{\rm d}}
\newcommand {\ee}{{\rm e}}

\newcommand {\bfa} {{\bf a}}
\newcommand {\bfk} {{\bf k}}

\newcommand {\bfp} {{\bf p}}
\newcommand {\bfr} {{\bf r}}
\newcommand {\bfv} {{\bf v}}

\newcommand {\bfR} {{\bf R}}

\newcommand {\calA} {{\cal A}}

\newcommand {\E}  {{\varepsilon}}
\newcommand {\om}  {{\omega}}
\newcommand {\Om}  {{\Omega}}

\newcommand{\Lch}{{L_{\rm ch}}}
\newcommand{\Ld}{{L_{\rm d}}}

\newcommand{\aTF}{{a_{\rm TF}}}

\newcommand{\MBNExplorer}{\textsc{MBN Explorer}\xspace}

\begin{document}

\title{
Channeling and Radiation of 855 MeV Electrons and Positrons in Straight and Bent 
Tungsten (110) Crystals}


%
\author{H. Shen} 
\affiliation{The Key Laboratory of Beam Technology and Material Modification of Ministry of Education, 
             College of Nuclear Science and Technology, 
             Beijing Normal University, Beijing 100875, China}
\affiliation{Beijing Radiation Center, Beijing 100875, China}
\affiliation{Center for Fusion Energy Science and Technology, Chinese Academy of 
             Engineering Physics, Beijing 100088, China}
\author{Q. Zhao}  
\affiliation{School of Nuclear Science and Engineering, North China Electric Power University, 
             Beijing 102206, China}

\author{F.S. Zhang  \footnote{Email: fszhang@bnu.edu.cn}}
\affiliation{The Key Laboratory of Beam Technology and Material Modification of Ministry of Education,
             College of Nuclear Science and Technology, 
             Beijing Normal University, Beijing 100875, China}
\affiliation{Beijing Radiation Center, Beijing 100875, China}
\affiliation{Center of Theoretical Nuclear Physics, National Laboratory of Heavy Ion Accelerator 
             of Lanzhou, Lanzhou 730000, China}
\author{Gennady B. Sushko} 
\affiliation{MBN Research Center, Altenh\"{o}ferallee 3, 60438 Frankfurt am Main, Germany}
\author{Andrei~V.~Korol 
\footnote{korol@mbnexplorer.com;
On leave from State Maritime University, St. Petersburg, Russia}
}
\affiliation{MBN Research Center, Altenh\"{o}ferallee 3, 60438 Frankfurt am Main, Germany}

\author{Andrey V. Solov'yov 
\footnote{solovyov@mbnresearch.com;
On leave from Ioffe Physical-Technical Institute,  St. Petersburg, Russia}
}
\affiliation{MBN Research Center, Altenh\"{o}ferallee 3, 60438 Frankfurt am Main, Germany}



\begin{abstract}
Planar channeling of 855 MeV electrons and positrons in straight and bent tungsten (110) 
crystal is simulated by means of the \MBNExplorer software package. 
The results of simulations for a broad range of bending radii 
are analyzed in terms of the channel acceptance, 
dechanneling length, and spectral distribution of the emitted radiation.
Comparison of the results with predictions of other theories as well as with
the data for (110) oriented diamond, silicon and germanium crystals is  carried out.  
\end{abstract}

\maketitle

\section{Introduction \label{Introduction}}

The basic effect of the channeling process in a straight crystal is in an anomalously 
large distance which a charged projectile can penetrate moving along a crystallographic 
direction (either planar or axial) experiencing collective action of the electrostatic field 
of the crystal atoms \cite{Lindhard1965}.
The channeling can also occur in a bent crystal provided the bending radius $R$ is 
large enough in comparison with the critical one $R_{\rm c}$ \cite{Tsyganov1976}.

A particle trapped into the channel oscillates in the transverse direction while propagating 
along a crystallographic direction.
The channeling oscillations give rise to a specific type of radiation, 
-- the channeling radiation (ChR) \cite{ChRad:Kumakhov1976}.
Its intensity depends on the type and the energy of a projectile as well as on the type 
of a crystal and a crystallographic direction.
The emission of ChR by an ultra-relativistic projectile in a straight crystal
is well studied (see, for example,  Refs.
\cite{Andersen_ChanRadReview_1983,BakEtal1985,Baier,BazylevZhevago,RelCha,%
Kumakhov2,Uggerhoj1993,Uggerhoj_RPM2005} and references therein).

The motion of a channeling particle in a bent crystal contains two components:
the channeling oscillations and circular motion along the bent centerline.
The latter motion gives rise to the synchrotron-type radiation (SR) \cite{Jackson}.  
Therefore, the total spectrum of radiation formed by an {ultra-relativistic} 
projectile in a bent crystal bears the features of both the ChR and SR.
Various aspects of radiation formed in crystals bent with the constant 
radius $R$ were discussed in Refs.
\cite{ArutyunovEtAl_NP_1991,Bashmakov1981,KaplinVorobev1978,%
SolovyovSchaeferGreiner1996,Taratin_Review_1998,TaratinVorobiev1988,TaratinVorobiev1989}
although with various degree of detail and numerical analysis.
A quantitative analysis of the emission spectra based on accurate simulation of the 
channeling process was carried out for silicon and diamond crystals
\cite{BentSilicon_2014,Sushko_EtAl_NIMB_v355_p39_2015,Polozkov_VKI_Sushko_AK_AS_SPB_Diamond_2015}.

The condition of stable channeling in a bent crystal, $R \gg R_{\mathrm{c}}$, 
\cite{Tsyganov1976} implies that the bending radius exceeds greatly the 
(typical) curvature radius of the channeling oscillations. 
Therefore, the SR modifies mainly the soft-photon part of the spectrum.
This part of the spectrum is especially interesting in connection with the concept  
of a crystalline undulator (CU) which implies propagation of ultra-relativistic projectiles 
along periodically bent crystallographic planes  \cite{KSG1998,KSG_review_1999}.
By means of CU it is feasible to produce undulator-like radiation 
in the hundreds of keV up to the MeV photon energy range. 
The intensity and characteristic frequencies of the radiation can 
be varied by changing the type of channeling particles,
the beam energy, the crystal type  and the parameters of periodic bending
(see recent review \cite{ChannelingBook2014}  for more details).
The range of bending amplitudes, $a$, and period, $\lambda_{\rm u}$, within which 
the operation of CU is feasible for projectiles of the energies $\E\simeq 10^{-1}-10^1$ GeV are:
$a\simeq10^0-10^1$ \AA, $\lambda_{\rm u} \simeq 10^0-10^2$ $\mu$m.
Even more exciting (although much more challenging and distant) is the possibility of generating 
a stimulated emission of the free-electron laser (FEL) type by means of CU 
\cite{KSG_review_1999,ChannelingBook2014}.
This novel source of light can generate, at least in theory, the stimulated emission
in the photon energy range $10^2-10^3$ keV (i.e. hard X and gamma range) which are not achievable 
in conventional FELs.

Up to now, several experiments have been (and planned to be) carried out to measure the 
channeling parameters and the radiation emitted by electrons and positrons
in periodically bent crystalline structures prepared by several different technologies 
\cite{ChannelingBook2014}.  
The recent attempts include experiments with 195--855 MeV electron beam at the Mainz Microtron 
(MAMI) facility
\cite{Backe_EtAl_2008,Backe_EtAl_2011,Backe_EtAl_2013,Backe_EtAl_PRL2014,BackeLauth_Dyson2016,%
BackeLauth:NIMB-v355-p24-2015}
carried out with CUs manufactured in Aarhus University (Denmark) using the molecular beam
epitaxy technology to produce strained-layer Si$_{1-x}$Ge$_{x}$ 
superlattices with varying germanium content \cite{MikkelsenUggerhoj2000}.
The CUs of this type are planned to be used in the coming experiments at SLAC 
with 10-35 GeV electron and positron FACET beam \cite{Uggerhoj2016}.
A set of experiments was performed with few GeV positrons at CERN \cite{Connell_Dyson2016}
with the CU based on synthetic diamond doped with boron \cite{DiamondBoron}. 
Other related experiments include investigations of the radiation of 
sub-GeV electrons in a bent silicon crystal \cite{Backe_EtAl_PRL_115_025504_2015}
and of the effectiveness  of deflection of multi-GeV electrons by a thin Si crystal
\cite{Wienands_EtAl_PRL_v114_074801_2015}. 

The technologies, available currently for preparing periodically bent crystals,
do not immediately allow for lowering the values of bending period down to 
tens of microns range or even smaller keeping, simultaneously, the bending 
amplitude in the range of several angstroms. 
These ranges of $a$ and $\lambda_{\rm u}$ are most favourable to achieve high intensity 
of radiation in a CU \cite{ChannelingBook2014}.
One of the potential options to lower the bending period is related to using 
crystals heavier than diamond ($Z=6$) and silicon ($Z=14$) to propagate ultra-relativistic 
electrons and positrons.
In heavier crystals, both the depth, $\Delta U\propto Z^{2/3}$, of the 
interplanar potential and its the maximum gradient, $U^{\prime}_{\max}\propto Z^{2/3}$, 
attain larger values, resulting
in the enhancement of the critical channeling angle $\Theta_{\rm L}\propto (\Delta U)^{1/2}$
\cite{Lindhard1965} and reduction of the critical radius $R_{\rm c} \propto 1/U^{\prime}_{\max}$ 
\cite{Tsyganov1976}.  
 
From this end, the tungsten crystal ($Z=74$) is a good candidate for the study.
This crystal was used in channeling experiments with both heavy 
\cite{Kovalenko_EtAl:JINR_RC-v72-p9-1995,BiryukovChesnokovKotovBook} 
and light \cite{Backe_EtAl:SPIE-6634-2007,Yoshida_EtAl:PRL_v80_p1437_1998}
ultra-relativistic projectiles. 
In Ref. \cite{Kovalenko_EtAl:JINR_RC-v72-p9-1995} it was noted that the 
straight tungsten crystals show high structure perfection. 
This feature is also of a great importance for successful experimental 
realization of the CU idea \cite{Imperfectness2008}. 
In the cited paper the comparison was carried out of the (110) planar channels in 
tungsten vs. silicon. 
In particular, the critical radius in W(110) was estimated as $R_{\mathrm{c}}=0.16$ cm which is 
 seven times smaller than that in Si(110).
This allows one, at least in theory, to consider periodic bending with 
$\lambda_{\rm u}\lesssim 10$ $\mu$m.
Indeed, matching the maximum curvature of periodic bending $4\pi^2a/\lambda_{\rm u}^2$
to $R_{\mathrm{c}}^{-1}$ one estimates $\lambda_{\rm u} \sim 2.5$ $\mu$m for $a=1$ \AA{}.

Therefore, it is desirable to carry out a quantitative analysis of both 
the channeling process and the emission of radiation of high-energy {\em light} projectiles  
in straight and bent tungsten crystal.
To the best of our knowledge, the simulations of these processes have been restricted to
the the straight oriented crystal and were carried out within model approaches.
In Refs. \cite{Azadegan_EtAl:JPConfSer-v517-p012039-2014,AzadeganWagner:NIMB-v517-p012039-2014} 
the continuous potential model was applied to construct the trajectories, whereas 
the dechanneling phenomenon was considered within the framework of the Fokker-Plank 
equation.
The model of binary collisions was exploited in Ref. \cite{Efremov_EtAl:RussPhysJ-v50-p1237-2007} 
to investigate the scattering angle of 0.5 GeV electrons and positrons 
in the process of axial channeling. 
No results have been presented for bent tungsten crystal. 

In this paper we present the results of simulation of the channeling process
and of the radiation emission for $\E=855$ MeV electrons and positrons in oriented
W(110) crystal.
The calculations were performed for both straight and bent crystals.
In the latter case the bending radius was varied down to 0.04 cm.
The projectile energy chosen corresponds to that used in the ongoing experiments 
with bent crystals and CUs at MAMI 
\cite{BackeLauth_Dyson2016,Backe_EtAl_PRL2014,BackeLauth:NIMB-v355-p24-2015} 
carried out with the electron beam.
However, from the view point of future experiments it is instructive to 
present a comparative analysis of the electron vs. positron channeling.
Therefore, both light projectiles are considered in the paper.

As in our recent studies, three-dimensional simulations of the 
propagation of ultra-relativistic projectiles through the crystal were performed 
by using the \MBNExplorer package \cite{MBN_ExplorerPaper,MBN_ExplorerSite}. 
The package was originally developed as a universal computer program to allow 
investigation of structure and dynamics of molecular systems of different origin 
on spatial scales ranging from nanometers and beyond. 
The general and universal design of the \MBNExplorer code made it possible to expand 
its basic functionality with introducing a module that treats classical relativistic
equations of motion and generates the crystalline environment
dynamically in the course of particle propagation \cite{ChanModuleMBN_2013}. 
A variety of interatomic potentials implemented in \MBNExplorer support 
rigorous simulations of various media. 
The software package can be regarded as a powerful numerical tool to uncover the dynamics 
of relativistic projectiles in crystals, amorphous bodies, as well as in biological environments. 
Its efficiency and reliability has already been benchmarked for the channeling of 
ultra-relativistic projectiles (within the sub-GeV to tens of GeV energy range) 
in straight, bent and periodically bent crystals 
\cite{ChanModuleMBN_2013,BentSilicon_2013,Sub_GeV_2013,BentSilicon_2014,SushkoThesis2015,%
Multi_GeV_2014,Sushko_AK_AS_SPB_SASP_2015,Sushko_EtAl_NIMB_v355_p39_2015,Si110-SASP-855MeV_2016}. 
In these papers verification of the code against available experimental
data and predictions of other theoretical models was carried out.

The description of the simulation procedure is sketched in Sect. \ref{Methodology}.
The results of calculations are presented and discussed in Sect. \ref{Results}.

\section{Methodology \label{Methodology}}

Propagation of an ultra-relativistic projectile of the charge $q$ and mass $m$ 
through a crystalline medium can be described in terms of classical relativistic dynamics.
This framework implies integration of the following two coupled equations of motion:
\begin{eqnarray}
\partial \bfr / \partial t = \bfv,
\qquad
\partial \bfp / \partial t = - q \, \partial U/\partial \bfr
\label{Methodology:eq.01} 
\end{eqnarray}
where $U=U(\bfr)$ is the electrostatic potential due to the crystal constituents, 
$\bfr(t), \bfv(t)$, and $\bfp(t) = m\gamma\bfv(t)$ stand, respectively, 
for the coordinate, velocity and momentum of the particle at instant $t$,   
$\gamma = \left(1-v^2/c^2\right)^{-1/2} = \E/mc^2$ is the relativistic Lorentz factor, 
$\E$ is the particle's energy, and $c$ is the speed of light. 

In \MBNExplorer, the differential equations (\ref{Methodology:eq.01}) are integrated using 
the forth-order Runge-Kutta scheme  with variable time step. 
At each integration step, the potential $U=U({\bf r})$ is calculated as the sum of atomic 
potentials $U_{\mathrm{at}}$ due to the atoms located inside the sphere of the cut-off radius 
$\rho_{\max}$ with the center at the instant location of the projectile. 
The value $\rho_{\max}$ is chosen large enough to ensure negligible contribution
to the sum from the distant atoms located at $r>\rho_{\max}$. 
The search for such atoms is facilitated by using the linked cell algorithm 
implemented in \MBNExplorer \cite{MBN_ExplorerPaper}.
The algorithm implies (i) a subdivision of the sample into cubic cells of a
 smaller size, and (ii) an assignment of each atom to a certain cell.
As a result, the total number of computational operations is reduced considerably.

To simulate the motion along a particular crystallographic plane with the Miller indices 
$(klm)$ the following algorithm is used \cite{ChanModuleMBN_2013}.

As a first step, a crystalline lattice is generated inside the rectangular simulation box
of the size $L_x\times L_y \times L_z$.
The $z$-axis is oriented along the beam direction and is parallel to
the $(klm)$ plane, the $y$ axis is perpendicular to the plane.
The position vectors of the nodes $\bfR_j^{(0)}$ ($j=1,2,\dots, N$) 
within the simulation box are generated in accordance with the 
type of the Bravais cell of the crystal and using the pre-defined values of 
the lattice vectors.

\MBNExplorer contains several build-in options which allow further modification
of generated crystalline structures.
The options relevant to modeling linear and bent crystals are as follows \cite{SushkoThesis2015}.

\begin{itemize}
 \item
\textit{Rotation} of the sample around a specified axis.
This option allows one to simulate the crystalline structure oriented along
any desired crystallographic direction.
In particular, this option allows one to choose the direction of the $z$-axis well away
from major crystallographic axes, thus avoiding the axial channeling (when not desired).

 \item
Displacement of the nodes in the transverse direction $y$ with respect to
a specified $z$-axis:
$y \to y + R(1-\cos\phi)$ where $\phi = \sin(z/R)$.
As a result, one obtains the crystalline structure
\textit{bent with constant radius} $R$ in the $(yz)$ plane.
For values of $R$  much larger than the crystal thickness $L$ (along
the $z$ direction), the displacement acquires the form:
\begin{equation}
y \to y + {z^2 \over 2R} \,.
\label{Methodology:eq.02}
\end{equation}

 \item
The nodes determine positions of the atoms in an ideal crystal.
More realistic structure includes the probability of the atoms to be
displaced from their equilibrium positions (the nodes) due to \textit{thermal vibrations}
corresponding to the given temperature $T$.
Thus, for each atom, the Cartesian components of the displacement are selected randomly 
by means of the normal distribution corresponding to fixed root-mean-square
amplitude $u_{T}$ of thermal vibrations. 
The values of $u_T$ for a number of crystals are summarized in \cite{Gemmel}.
For tungsten one finds $u_T=0.05$ \AA.

Let us note that by introducing unrealistically large value of $u_T$
(for example, exceeding the lattice constants)
it is possible to consider large random displacements.
As a result, the \textit{amorphous medium} can be generated.

\item
\textit{Periodic harmonic displacement of the nodes} can be performed
 by means of the transformation
 $\bfr \to  \bfr + \bfa \sin(\bfk\cdot\bfr+\varphi)$.
 Here, the vector $\bfa$ and its modulus, $a$, determine the direction and the amplitude of 
 the displacement,
 the wave-vector $\bfk$ determines the axis along which the displacement to be propagated,
 and $\lambda_{\rm }=2\pi/k$ defines the wave-length of the periodic bending.
The parameter $\varphi$ allows one to change the phase-shift of the harmonic bending.

 In a special case $\bfa \perp \bfk$, this options provides simulation of
 linearly polarized periodically bent crystalline structure which is 
 an important element of a crystalline undulator.

\end{itemize}

In addition to the aforementioned options, \MBNExplorer allows one to 
model periodically bent crystalline structures by generating  
periodic (harmonic) displacement of the nodes,
to construct binary structures (for example, Si$_{1-x}$Ge$_x$ superlattices)
by introducing random or regular substitution of atoms in the initial structure
with atoms of another type.
Also, the simulation box can be cut along specified faces, thus allowing tailoring the 
generated crystalline sample to achieve the desired form of the sample.

Trajectory of a particle entering the initially constructed crystal at the instant $t=0$
is calculated by integrating equations (\ref{Methodology:eq.01}).
Initial transverse coordinates, $(x_0, y_0)$, and velocities, $(v_{x,0}, v_{y,0})$, are 
generated randomly accounting for the conditions at the crystal entrance 
(i.e., the crystal orientation, beam emittance and energy distribution of the particles).
A particular feature of \MBNExplorer is in simulating the crystalline environment "on the fly", 
i.e. in the course of propagating the projectile.
This is achieved by introducing a dynamic simulation box which shifts following 
the particle (see Ref.~\cite{ChanModuleMBN_2013} for the details). 

Taking into account randomness in sampling the incoming projectiles and 
in positions of the lattice atoms due to the thermal fluctuations, 
one concludes that each simulated trajectory corresponds to a unique crystalline
environment.
Thus, all simulated trajectories are statistically independent and can be
analyzed further to quantify the channeling process as well as the emitted radiation.

The averaged spectral distribution of the energy emitted within the cone 
$\theta\leq \theta_{0}$ with respect to the incident beam (i.e. along the $z$ axis)
 is computed as follows 
\begin{eqnarray}
\left\langle{\d E(\theta\leq\theta_{0}) \over \d \om} \right\rangle
=
{1 \over N_0}
\sum_{n=1}^{N_0} 
\int\limits_{0}^{2\pi}
\d \phi
\int\limits_{0}^{\theta_{0}}
\theta \d\theta\,
{\d^2 E_n \over \d \om\, \d\Om}.
\label{Methodology:eq.03}
\end{eqnarray}
Here, $\om$ stands for the frequency of radiation, 
$\Omega$ is the solid angle corresponding to the emission angles $\theta$ and $\phi$.
The sum is carried over the simulated trajectories of the total number $N_0$, 
and $\d^2 E_n/\d \om\, \d\Om$ is the energy per unit frequency and unit solid angle 
emitted by the projectile moving along the $n$th trajectory. 
The resulting spectrum accounts for all mechanisms of the radiation formation:
(a) channeling radiation (ChR) due to the channeling segments,
(b) coherent and incoherent bremsstrahlung (BrS) due to the over-barrier motion.
In addition to these, the motion along the arc in a bent crystal results in the
synchrotron-type radiation. 

The numerical procedures implemented in \MBNExplorer to calculate the distributions 
$\d^2 E_n/\d \om\, \d\Om$ \cite{ChanModuleMBN_2013}
are based on the quasi-classical formalism due to Baier and Katkov \cite{Baier}.
A remarkable feature of this method is that it combines classical description of the 
motion in an external field with the quantum corrections due to the 
radiative recoil quantified by the ratio $\hbar\om/\E$.
In the limit  $\hbar \om /\E \ll 1$ one can use the classical
description of the radiative process which is adequate to describe the 
emission spectra by electrons and positrons of the sub-GeV energy range 
(see, for example,~\cite{ChannelingBook2014} and references therein).  
The corrections lead to strong modifications of
the radiation spectra of multi-GeV projectiles channeling in crystalline undulators
\cite{Multi_GeV_2014,Sushko_AK_AS_SPB_SASP_2015} and in bent crystals 
\cite{Sushko_EtAl_NIMB_v355_p39_2015}.

\section{Results and Discussion \label{Results}}

Using the algorithm outlined above, classical trajectories were simulated for 
$\E=855$ MeV electrons and positrons incident along the (110) crystallographic
plane in straight and bent tungsten crystals of the thickness $L_1=75$ $\mu$m along the
incident beam direction.

In a bent crystal, the channeling condition \cite{Tsyganov1976} implies that 
the centrifugal force $F_{\rm cf}= pv/R \approx \E/R$ is smaller than the maximum 
interplanar force $F_{\max}$.
It is convenient to quantify this statement by introducing the dimensionless 
bending parameter $C$:
\begin{equation}
C = {F_{\rm cf} \over F_{\max}} = {\E \over R F_{\max}} = {R_{\rm c} \over R}.
\label{Results:eq.01}
\end{equation}
The case $C = 0$ ($R=\infty$) characterizes the straight crystal whereas
$C = 1$ corresponds to Tsyganov's critical (minimum) radius $R_{\rm c}=\E/F_{\max}$
\cite{Tsyganov1976}. 
Within the framework of the continuous interplanar potential model \cite{Lindhard1965},
one calculates $F_{\max}=42.9$ GeV/cm for W(110) at room temperature 
by means of the formula for the continuous potential derived in Ref. 
\cite{Erginsoy_PRL_v15_360_1965,AppletonEtAl_PR_v161_330_1967} based on the Moli\`ere 
approximation for the atomic potential \cite{Moliere}.
Hence, $R_{\rm c}\approx0.02$ cm for a 855 MeV projectile. 

In the calculations presented below the bending radius was varied from 2 down to
0.0432 cm.
For each type of the projectiles and for each value of bending radius 
(including the case of the straight crystal), the numbers $N_0$ of the 
simulated trajectories were sufficiently large (approximately $5000$), thus enabling 
a reliable statistical quantification of the channeling process. 
In Sect. \ref{Trajectories} below we define and describe the quantities obtained. 
The emission spectra are discussed in Sect. \ref{Spectra}.

\subsection{Statistical Analysis of Trajectories \label{Trajectories}} 

\begin{figure}
\includegraphics[width=13.0cm,clip]{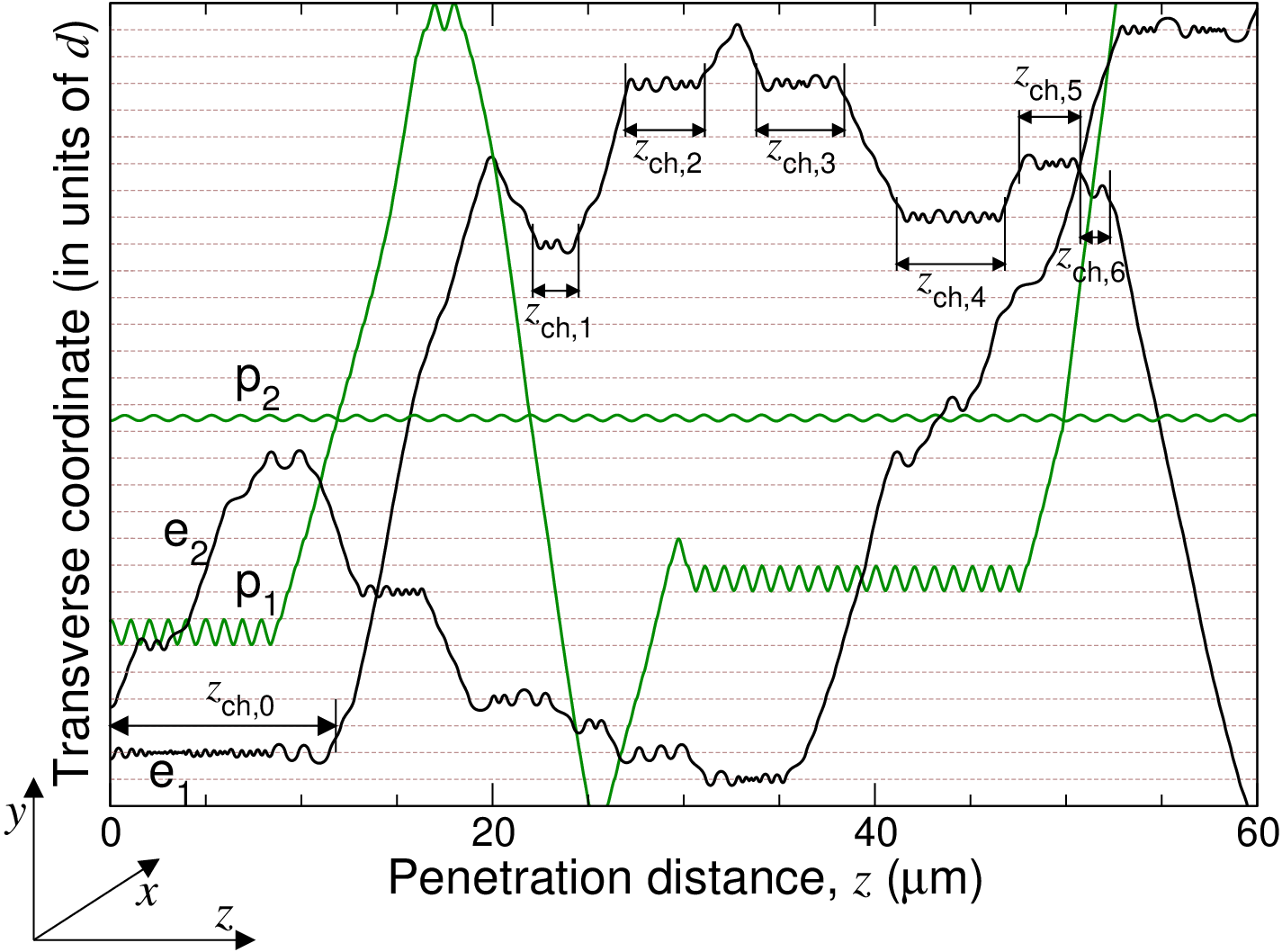}
\caption{
Selected simulated trajectories of 855 MeV electrons (curves "$\ee_1$" and "$\ee_2$") 
and of positrons (curves "${\rm p}_1$" and "${\rm p}_2$") 
propagating in straight W(110).
The trajectories illustrate the channeling and the over-barrier motion as well
as the dechanneling and rechanneling effects. 
The $z$-axis of the reference frame is directed along the incoming projectiles, the 
$(xz)$-plane is parallel to the (110) crystallographic planes (dashed lines) 
and the $y$-axis is perpendicular to the planes. 
The W(110) interplanar distance is $d=2.238$ \AA.
In the electron case, presented are the accepted ("$\ee_1$") and non-accepted ("$\ee_2$")
trajectories. 
For the accepted trajectory the characteristic lengths of the channeling motion are 
indicated: the initial channeling segment, $z_{{\rm ch}0}$, and 
the segments in the bulk, $z_{{\rm ch}1}, \dots, z_{{\rm ch}6}$.
The non-accepted trajectory corresponds to $z_{{\rm ch}0} = 0$.
The positron trajectories both refer to the accepted type.
}
\label{Figure.01} 
\end{figure}

Fig. \ref{Figure.01} shows simulated trajectories of 855 MeV electrons (two black curves labeled 
"$\ee_1$" and "$\ee_2$") and of positrons (two green curves "${\rm p}_1$" and "${\rm p}_2$") 
propagating in straight W(110) crystal.
These selected trajectories as well as other notations and features presented in the figure
we use in the text below as a reference material when explaining various quantitative 
and qualitative characteristics of the particles motion.

To start with, we explain the geometrical parameters used in the simulations.
Dashed horizontal lines in Fig. \ref{Figure.01} mark the cross section of the (110) 
crystallographic planes separated by the distance $d=2.238$ \AA{}.
Thus, the $y$-axis is aligned with the $\langle 110 \rangle$ crystallographic axis.
The horizontal $z$-axis corresponds to the direction of the incoming beam which was considered
 ideally collimated in the current simulations.
To avoid the axial channeling, the $z$-axis was chosen along the 
$[10, -10, 1]$ crystallographic direction.
A projectile enters the crystal at $z=0$ and exits at $z=L$.
The crystal is considered infinitely large in the $x$ and $y$ directions.
In the simulations, the integration of the equations of motion (\ref{Methodology:eq.01})
produces a 3D trajectory. 
The black and green curves in the figure represent the projections of the corresponding 
3D trajectories on the $(yz)$ plane.  

The trajectories shown in Fig. \ref{Figure.01} illustrate a variety of features 
which characterize the motion of a charged projectile in an oriented crystal: 
the channeling mode, the over-barrier motion, the dechanneling and the rechanneling processes,
rare events of hard collisions etc. 

The channeling motion is more pronounced and regular for positrons than for electrons.
Positively charged projectiles tend to move in between the atomic planes, i.e.
in the region with lower volume density of the crystal constituents. 
As a result, they tend to stay in the channeling mode much longer than negatively charged particles,
which move in the vicinity of the atomic chains being attracted by positively charged 
nuclei.
On the other hand, although the electrons have higher rate of the dechanneling
(i.e., leaving the channeling mode of motion due to the collisions), the inverse phenomenon,
the re-channeling, is also more frequent for them \cite{Kostyuk-AK-AS-WG_855-Si}.
Therefore, as it is seen in the figure, quite often the electron trajectory consists of several
channeling segments separated by the intervals of the over-barrier motion.  
The nearly harmonic pattern of the channeling oscillations for positrons 
as well as strong anharmonicity in the electron channeling oscillations 
are well-known phenomena (see, e.g., \cite{BakEtal1985}).
These effects can be explained qualitatively within the framework of the continuous 
potential model by comparing the profiles of the interplanar potentials 
for the two types of projectiles (nearly harmonic for positrons and extremely non-harmonic 
for electrons).  

Apart from providing the possibility of illustrative comparison, the simulated trajectories
allow one to quantify the channeling process in terms of several parameters and 
functional dependencies which can be generated on the basis of statistical
analysis of the trajectories 
\cite{ChanModuleMBN_2013,ChannelingBook2014,BentSilicon_2013,Sub_GeV_2013,BentSilicon_2014,
Sushko_EtAl_NIMB_v355_p39_2015,Sushko_AK_AS_SPB_SASP_2015,Si110-SASP-855MeV_2016}.

A randomization of the "entrance conditions" for the projectiles, 
as explained in Sect.~\ref{Methodology}, leads to different chain of scattering 
events the different projectiles at the entrance to the bulk. 
As a result, not all the simulated trajectories start with the channeling segments.
In Fig. \ref{Figure.01},  trajectory "$\ee_2$" refers to the non-accepted projectile while three 
other trajectories correspond to the accepted ones.
A commonly used parameter to quantify this feature is 
{\em acceptance} defined as the ratio $\calA = N_{\rm acc}/N_0$ of the number 
$N_{\rm acc}$ of particles captured into the channeling mode at the entrance  
of the crystal (the accepted particles) to the total number $N_0$ of the incident particles. 
The non-accepted particles experience unrestricted over-barrier motion at the entrance 
but can rechannel somewhere in the bulk.

\begin{figure}
\includegraphics[width=13.0cm,clip]{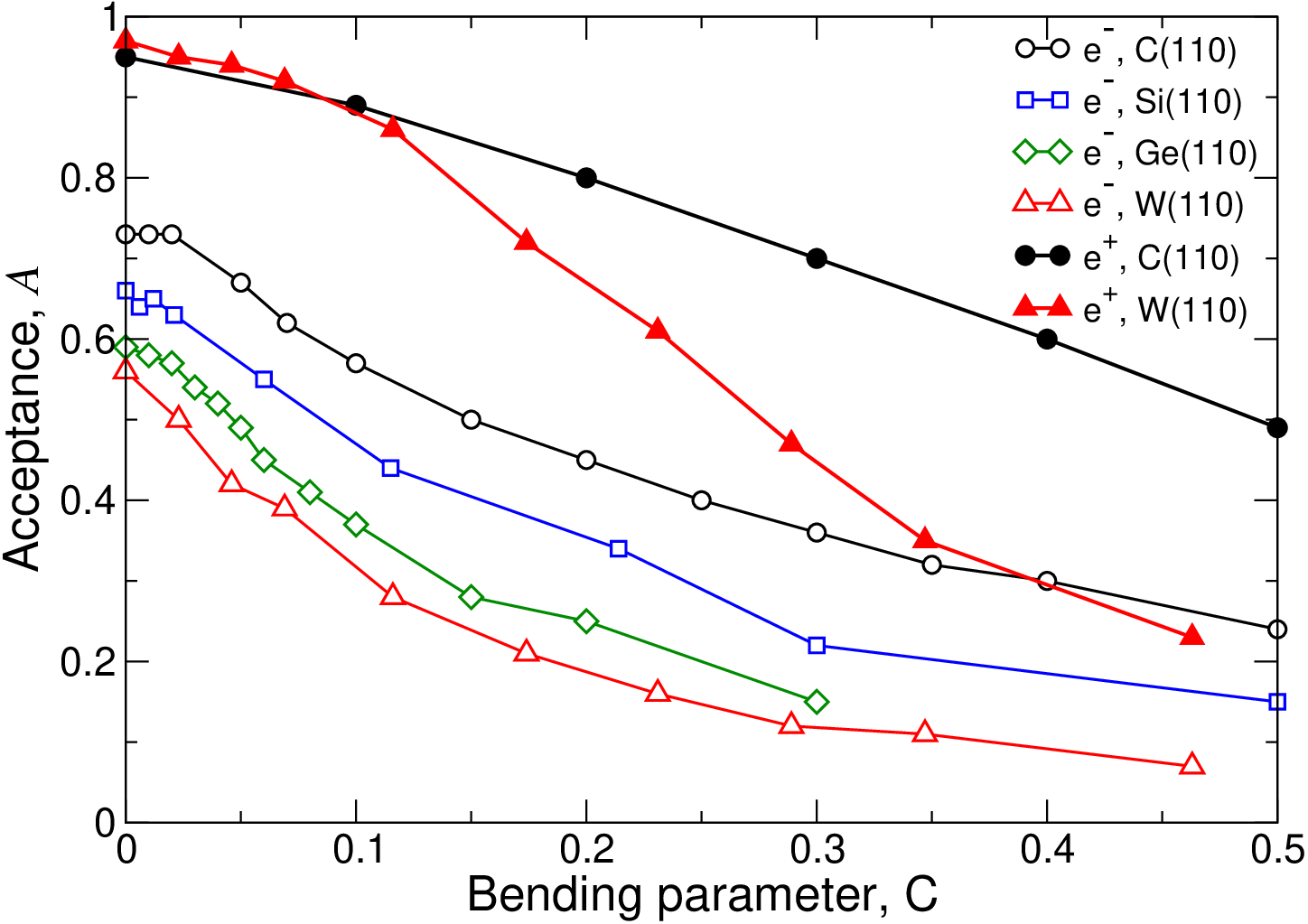}
\caption{
Acceptance versus bending parameter $C$ (see Eq. (\ref{Results:eq.01}))
for 855 MeV electrons (open symbols) and positrons (filled symbols).
The results of current simulations for W(110) are compared with the 
data generated previously by means of \MBNExplorer:
diamond(110) from Ref. \cite{Sushko_AK_AS_SPB_SASP_2015},
Si(110) from Refs. \cite{BentSilicon_2014,BentSilicon_2013},
and Ge(110) from Ref. \cite{SushkoThesis2015}.
}
\label{Figure.02} 
\end{figure}

The acceptance as a function of the bending parameter $C$ is presented
in Fig. \ref{Figure.02}. 
The results of the current set of simulations for W(110) are plotted together with the data 
obtained earlier for other oriented crystals, as indicated in the caption.
For each crystal, the corresponding values of bending radius $R$
are calculated from Eq. (\ref{Results:eq.01}) using the following values of the (110) 
interplanar field $F_{\max}$ at room temerature:
7.0, 5.7, 10.0, and 42.9 GeV/cm for diamond, silicon, germanium, and tungsten, 
respectively.
To be noted is 
(i) the monotonous decrease of $\calA$ with $C$,
(ii) larger acceptances for positrons than for electrons for given $C$ and crystal,
and
(iii) non-monotonous behaviour of  $\calA$ with respect to the charge number $Z$
of the crystal atom (in the positron case).
Feature (i) is readily explained using the concept of the continuous interplanar 
potential and its modification in the bent channel 
(decrease of the effective potential due to the centrifugal term which increases with $C$)  
\cite{BakEtal1985,Ellison:NP_B-v206-p205-1982,BiryukovChesnokovKotovBook,Tabrizi_AK_AS_WG:JPG}. 
Feature (ii) follows from the fact that in an oriented crystal a projectile electron, being 
attracted to an atomic chain (plane), gains the transverse energy via the collisions with the 
crystal constituents and, thus, switches to the over-barrier motion 
much faster than a positron.  
Feature (iii) reflects the resulting impact of the two opposite tendencies.
First, both the average value of the interplanar field and the interplanar spacing 
increase with $Z$, thus acting towards the increase of the acceptance.
The opposite tendency, is the increase of the (average) scattering angles in collisional 
events of the projectile with heavier atoms.
For electrons, which tend to move in the vicinity of the atomic chains, 
the latter tendency outpowers the former one.
For positrons, the two curves presented (for $Z=6$ and $Z=74$) indicate that dependence 
$\calA$ on $Z$ is not so strong as for electrons.
We note that in Ref. \cite{Kovalenko_EtAl:JINR_RC-v72-p9-1995} the acceptance of W(110) 
was estimated to be slightly higher than for Si(110).

Within the framework of the continuous potential approximation \cite{Lindhard1965},
the criterion for distinguishing between the channeling and non-channeling motions 
is introduced straightforwardly by matching the value of the transverse energy, $\E_{\perp}$,
of the projectile with the depth, $\Delta U$, of the interplanar potential well.
The channeling mode corresponds to $\E_{\perp} < \Delta U$. 
In the simulations based on solving the equations of motion (\ref{Methodology:eq.01})
with the potential built up as an exact sum of atomic potentials,
another criterion is required to select the channeling segments.
Following  \cite{ChanModuleMBN_2013}, we assume the channeling to occur when a projectile, 
while moving in the same channel, changes the sign of the transverse 
velocity $v_y$ at least two times.

An accepted projectile, stays in the channeling mode of motion over some interval
$z_{{\rm ch},0}$ until an event of the dechanneling (if it happens).
The initial channeling segment $z_{{\rm ch},0}$ is explicitly indicated for trajectory
"$\ee_1$" in Fig. \ref{Figure.01}.
Trajectory "$\ee_2$" corresponds to the non-accepted particle ($z_{{\rm ch},0}=0$).
Both positron trajectories stand for the accepted particles, and in case of "${\rm p}_2$"
the dechanneling does not occur within the indicated length of the crystal. 
To quantify the dechanneling effect for the accepted particles, one can introduce 
{\em the penetration depth} $L_{\mathrm{p1}}$ \cite{ChanModuleMBN_2013} 
defined as the arithmetic mean of the initial channeling segments $z_{{\rm ch},0}$
calculated with respect to all accepted trajectories.
For sufficiently thick crystals ($L \gg L_{\mathrm{p1}}$), the penetration depth 
approaches the so-called dechanneling length $\Ld$ which characterizes 
the fraction of the channeling particles at large distances $z$ from the entrance 
in terms of the exponential decay, $\propto \exp(-z/\Ld)$. 
(see, e.g., Ref.~\cite{BiryukovChesnokovKotovBook}).
We note here that the concept of the exponential decay has been 
widely exploited to estimate the de-channeling lengths for various 
ultra-relativistic projectiles in the straight and bent crystals 
\cite{Backe_EtAl_2008,BogdanovDabagov2012,Scandale_EtAl_2012,BiryukovChesnokovKotovBook,
PhysRevLett.112.135503,Wienands_EtAl_PRL_v114_074801_2015}.

Random scattering of the projectiles on the crystal atoms can result in the {\em rechanneling}, 
i.e. the process of capturing the over-barrier particles into the channeling mode of motion. 
In a sufficiently long crystal, the projectiles can experience dechanneling and 
rechanneling several times in the course of propagation. 
These multiple events can be quantified by introducing the penetration length 
$L_{\mathrm p2}$ and the total channeling length $\Lch$.
These quantities characterize the channeling process in the whole crystal. 
In explaining the meaning of the depth $L_{\mathrm p2}$ we refer to the electron trajectories
 in Fig. \ref{Figure.01}.
Trajectory "$\ee_1$" has a non-zero initial channeling segment, $z_{{\rm ch},0}$, as well as
several secondary channeling segments, marked as $z_{{\rm ch}1}, \dots, z_{{\rm ch}6}$, which
appear due to the rechanneling process. 
So, altogether this trajectory contains seven channeling segments. 
Trajectory "$\ee_2$" exhibit only secondary channeling segments (not marked explicitly) of the
total number ten.
The penetration length $L_{\rm p2}$ is calculated as the arithmetic mean of all channeling 
segments (initial and secondary) with respect to the total number of channeling segments in all 
simulated trajectories.
Finally, the total channeling length $\Lch$ per particle is calculated by averaging the sums 
$z_{{\rm ch}0} + z_{{\rm ch}1} + z_{{\rm ch}2} + \dots$, calculated for each trajectory, 
over all trajectories. 

To conclude the description of the lengths introduced above, we note that $L_{\rm p1}$ 
characterizes the distance covered by accepted particles moving in the channeling mode.
Generally speaking, it depends on the beam emittance at the crystal entrance.
The second penetration depth, $L_{\rm p2}$, accounts for the rechanneling events, which 
occur, on average, for the incident angles not greater than Lindhard's critical angle 
$\Theta_{\rm L}$ \cite{Lindhard1965}.
Hence, for sufficiently long crystals $L_{\rm p2}$ mimics the initial penetration depth
of the beam with a non-zero emittance equal approximately to  
$\Theta_{\rm L}=(2\Delta U_0/\E)^{1/2} \approx 0.56$ mrad.
The latter estimate was obtained for a $\E=855$ MeV projectile using the value 
$\Delta U_0 = 132.2$ eV for the interplanar potential well in straight W(110) channel
at room temperature.

The results on the acceptance and the characteristic lengths are presented in 
Table~\ref{Table:ep-A-Lp12-Lch}.  
All the data refer to zero emittance beams entering a $L=75$ $\mu$m W(110) crystal. 
The straight crystal corresponds to infinitely large bending radius, $R=\infty$.
Statistical uncertainties due to the finite numbers $N_0$ of the simulated trajectories correspond 
to the $99.9 \%$ confidence interval.

\begingroup
\squeezetable
\begin{table}
\caption{Acceptance $\calA$, bending parameter $C$,
penetration lengths $L_{\rm p1,2}$ and total channeling $\Lch$ length
(all in $\mu$m) for 855 MeV electrons and positrons in straight ($R=\infty$) and 
bent ($R<\infty$, in cm) W(110) crystal.
\label{Table:ep-A-Lp12-Lch}
}
\begin{ruledtabular}
\begin{tabular}{cc|cccc|cccc}
         &       &   \multicolumn{4}{c|}{Electron}&   \multicolumn{4}{c}{Positron}\\
  $R$    & $C$   
                 &$\calA$& $L_{\rm p1}$  & $L_{\rm p2}$   & $\Lch$ 
                 &$\calA$& $L_{\rm p1}$  & $L_{\rm p2}$   & $\Lch$\\
\hline
$\infty$&$0.00$ &$0.56$& $3.33\pm0.13$& $3.43\pm 0.05$& $12.7 \pm 0.5$ 
                &$0.97$& $69.4\pm 1.0$& $59.0\pm 1.6 $& $68.1 \pm 1.0 $\\
$2.0$   &$0.01$ &$0.53$& $3.19\pm0.12$& $3.38\pm 0.05$& $8.85 \pm 0.37$ 
                &      &              &               &  \\
$0.864$ &$0.023$&$0.50$& $3.16\pm0.11$& $3.29\pm 0.07$& $4.13 \pm 0.18$ 
                &$0.95$& $68.8\pm 0.9$& $66.7 \pm 1.0$& $65.8 \pm 0.9$ \\
$0.432$ &$0.046$&$0.42$& $2.90\pm0.14$& $2.94\pm 0.11$& $1.94 \pm 0.14$ 
                &$0.94$& $68.3\pm 0.9$& $67.4\pm 1.0 $& $64.4 \pm 0.9$ \\
$0.288$ &$0.069$&$0.39$& $2.64\pm0.13$& $2.66\pm 0.11$& $1.32 \pm 0.11$ 
                &$0.92$& $66.4\pm 1.0$& $65.9\pm 1.0 $& $61.4 \pm 1.0$  \\
$0.173$ &$0.116$&$0.28$& $2.23\pm0.12$& $2.24\pm 0.11$& $0.71 \pm 0.08$ 
                &$0.86$& $61.9\pm 1.4$& $61.6\pm 1.4 $& $53.1 \pm 1.4$ \\
$0.115$ &$0.174$&$0.21$& $1.79\pm0.09$& $1.80\pm 0.09$& $0.40 \pm 0.05$ 
                &$0.72$& $55.2\pm 1.8$& $55.0\pm 1.8 $& $39.8 \pm 1.7$ \\
$0.086$ &$0.231$&$0.16$& $1.58\pm0.09$& $1.58\pm 0.09$& $0.26 \pm 0.04$ 
                &$0.61$& $49.8\pm 2.2$& $49.5 \pm 2.2$& $30.2 \pm 1.9 $\\
$0.069$ &$0.289$&$0.12$& $1.34\pm0.09$& $1.34\pm 0.09$& $0.17 \pm 0.0$ 
                &$0.47$& $45.6\pm 2.5$& $45.5 \pm 2.5$& $21.5 \pm 2.0 $\\
$0.058$ &$0.347$&$0.11$& $1.25\pm0.09$& $1.25\pm 0.09$& $0.14 \pm 0.03$ 
                &$0.35$& $38.2\pm 2.8$& $38.1 \pm 2.8$& $13.3 \pm 1.7$ \\
$0.043$ &$0.463$&$0.07$& $1.03\pm0.08$& $1.03\pm 0.08$& $0.07 \pm 0.02$ 
                &$0.23$& $18.5\pm 1.7$& $18.4 \pm 1.7$& $4.2 \pm 0.7$  \\
\end{tabular}
\end{ruledtabular}
\end{table}
\endgroup

We start discussion of the data presented in Table \ref{Table:ep-A-Lp12-Lch} with the 
penetration lengths of electrons. 

The length $L=75$ $\mu$m of the crystal exceeds greatly the quoted values of $L_{\rm p1}$ and 
$L_{\rm p2}$. 
Therefore, these quantities can be associated 
with the dechanneling lengths of the ideally collimated electron beam 
(the penetration depth $L_{\rm p1}$) and of 
the beam with emittance of approximately $\Theta_{\rm L}$ (the depth $L_{\rm p2}$).
For a collimated electron beam, to {\em estimate} the dechanneling length in straight W(110) 
channel  one can use its relationship \cite{Baier} to the radiation length
$L_{\rm r}$ in the corresponding amorphous medium.
This relationship can be written in the following form, which is convenient for
a quick estimate of  $\Ld$ (see Eq. (6.4) in Ref. \cite{ChannelingBook2014}):
\begin{equation}
\Ld \, \mbox{[$\mu$m]} 
= 0.089\times \Delta U_0 \mbox{[eV]}\, \E \mbox{[GeV]}\, L_{\rm r}  \mbox{[cm]}.
\label{Results:eq.02}
\end{equation}
For a 855 MeV electron in W(110) ($\Delta U_0 = 132.2$ eV, $L_{\rm r} =0.35$ cm) this
formula results in $\Ld = 3.5$ $\mu$m which correlates with values of 
$L_{\rm p1}$ and $L_{\rm p2}$ presented in the table for the straight channel ($R=\infty$).
We note that all these values are higher than the dechanneling length which 
follows from the consideration presented in Ref. \cite{Azadegan_EtAl:JPConfSer-v517-p012039-2014}
where electron dechanneling process in straight W(110) was analyzed in terms of 
the Fokker-Planck equation.
As a result of the analysis, the authors provide the following dependence of $\Ld$ (in $\mu$m) 
on the electron beam energy $\E$ (in GeV): $\Ld=2.78\E$.
For a 855 MeV projectile this results in $\Ld=2.4$ $\mu$m.

It is instructive to compare the estimates $\Ld$ as they follow from Eq. (\ref{Results:eq.02})
with the results for $L_{\rm p1}$ obtained previously for 855 MeV electrons by means 
of \MBNExplorer for other oriented crystals.
The simulated data are: 
$L_{\rm p1}=12.01 \pm 0.40$  $\mu$m for diamond(110) \cite{Sushko_AK_AS_SPB_SASP_2015},
$11.72 \pm 0.30$  $\mu$m for Si(110) \cite{BentSilicon_2014,BentSilicon_2013},
and 
$6.57 \pm 0.30$  $\mu$m for Ge(110) \cite{SushkoThesis2015}.
For the same crystals, the estimated values are 
$\Ld=18.1, 15.7,$ and 6.90 $\mu$m, respectively.
Hence, the observation is that Eq. (\ref{Results:eq.02}) overestimates the dechanneling
length for low-$Z$ crystals (diamond and silicon) but provides good results for 
medium- and high-$Z$ ones (germanium and tungsten).

The penetration lengths $L_{\rm p1,2}$ (as well as the dechanneling length) 
decrease with the increase in the bending parameter $C$.
Within the framework of the continuous potential model this feature can be explained in terms 
of the depth of the potential well which in a bent channel,  $\Delta U_C$, is smaller than
in a straight one, $\Delta U_0$, due to the centrifugal term (see, e.g., Refs.
\cite{Ellison:NP_B-v206-p205-1982,Tabrizi_AK_AS_WG:JPG,BiryukovChesnokovKotovBook}).

It is instructive to compare the values of $L_{\rm p1,2}$ with the total channeling 
length $\Lch$ of an electron.
In the straight channel, $\Lch$ exceeds $L_{\rm p1,2}$ approximately by a factor of four.
Thus, on average, an electron trajectory contains four channeling segments when propagating through 
a 75 $\mu$m thick W(110) oriented crystal.
For an accepted particle, one of these is the initial channeling segment whereas other three 
are due to the rechanneling.
For a non-accepted particle, all four segments appear as a result of the rechanneling events.
These figures allow one to estimate the {\em rechanneling} length, 
$L_{\rm rech} \approx 20$ $\mu$m, 
i.e. the average length of a segment within which a projectile moves in the over-barrier mode.
The table shows that as $C$ increases, the decrease rate of $\Lch$ is much larger than that 
of $L_{\rm p1,2}$.
This is a clear indication that in a bent crystal the rechanneling events are much rarer than 
in a straight one.
Starting with some bending parameter $\widetilde{C}$ the rechanneling events virtually cease 
to occur.
To estimate $\widetilde{C}$ one can compare the value of $N_{\rm acc} L_{\rm p1}/N_0 
=\calA  L_{\rm p1}$,
i.e. the initial channeling segments averaged over all trajectories, including the
non-accepted ones, with $\Lch$.
For $C\geq \widetilde{C}$ one obtains $\calA  L_{\rm p1}\approx \Lch$.
The electron data presented in Table \ref{Table:ep-A-Lp12-Lch} suggests that this approximate 
equality becomes valid starting with $C\approx 0.1$. 
Hence, in W(110) channel bent with radius $R\approx R_{\rm c}/10$ or smaller
the rechanneling event virtually do not happen (the corresponding rechanneling 
lengths become infinitely large).

\begin{figure}
\centering
\includegraphics[width=7.5cm,clip]{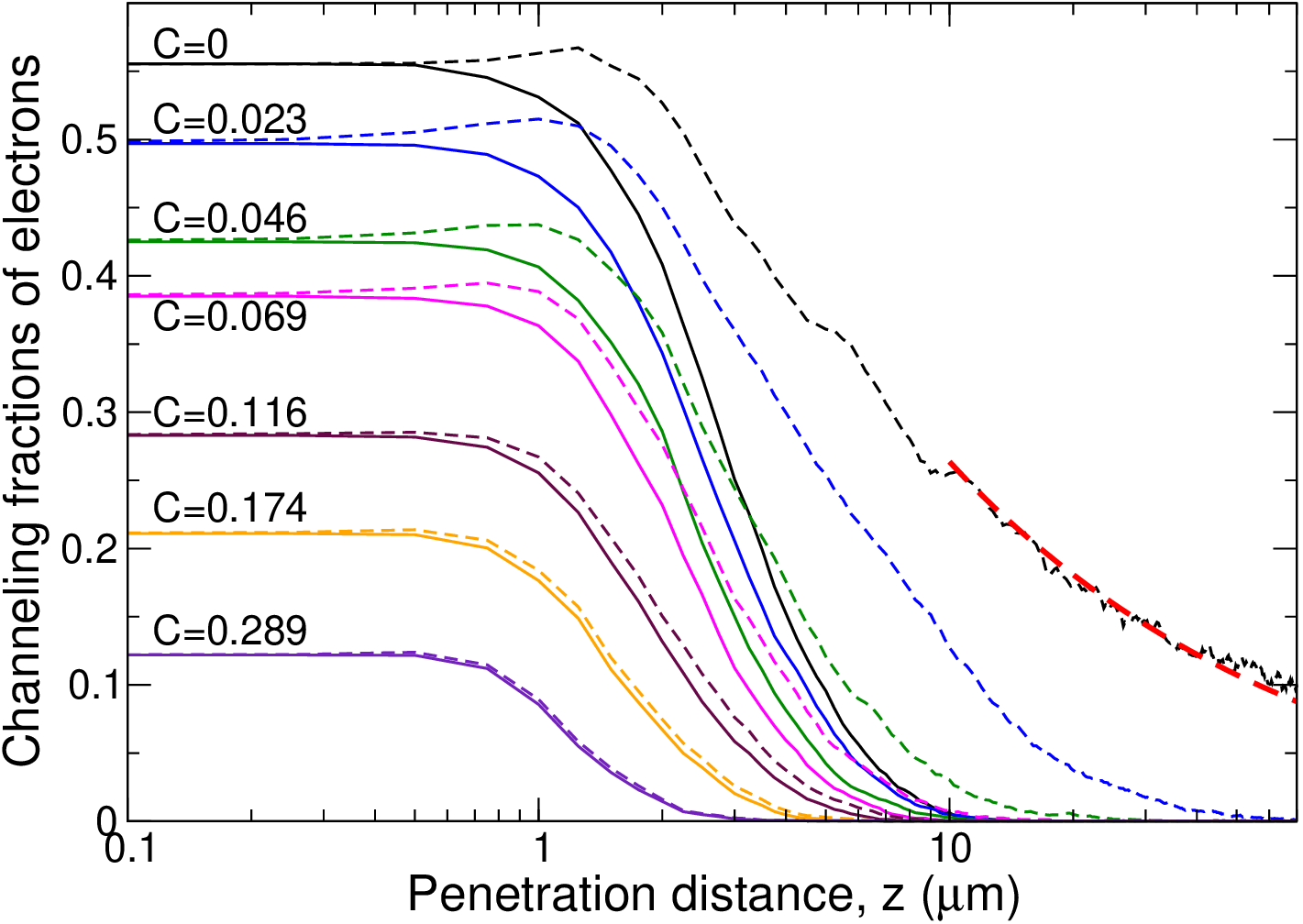}
\hspace*{0.5cm}
\includegraphics[width=7.5cm,clip]{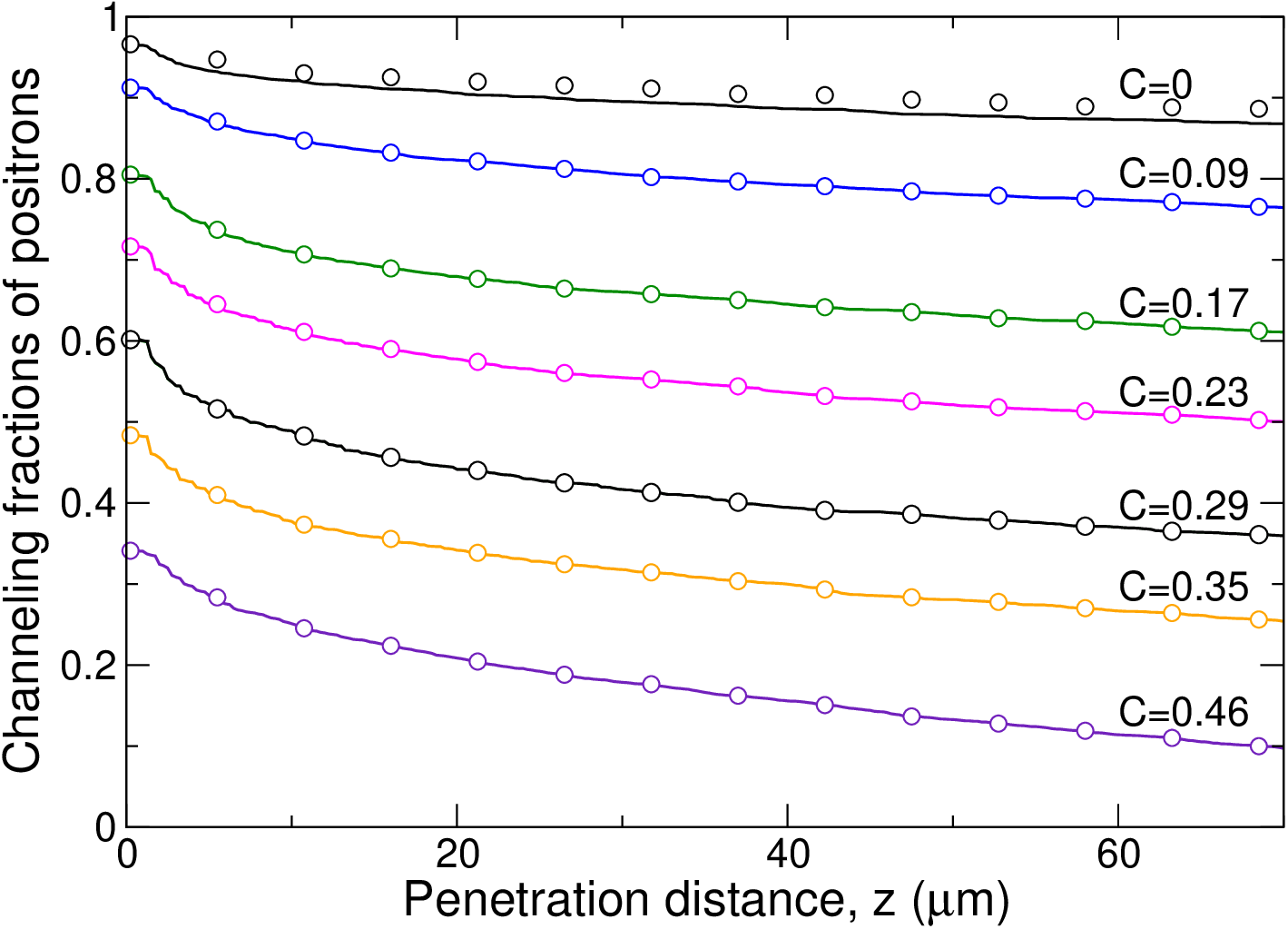}
\caption{
Channeling fractions $\xi_{\rm ch0}(z)$ (solid curves) and 
$\xi_{\rm ch}(z)$ (dashed curves in the left panel, symbols in the right panel) 
calculated for 855 MeV electrons (left panel, note the 
log scale of the horizontal axis)
and positrons (right panel) in straight ($C=0$) and bent ($C>0$) 
W(110) channels.
Thick (red) line in the left graph shows the dependence 
$\xi_{\rm ch}(z)\propto z^{-1/2}$, see explanation in the text.
}
\label{Figure.03} 
\end{figure}

To further quantify the impact of the rechanneling effect basing on the information which can 
be extracted from the simulated trajectories, one can compute the {\em channeling fractions}  
$\xi_{\rm ch0}(z)=N_{\rm ch0}(z)/N_0$ and $\xi_{\rm ch}(z)=N_{\rm ch}(z)/N_0$
\cite{ChanModuleMBN_2013,BentSilicon_2013}.
Here $N_{\rm ch0}(z)$ stands for the number of electrons (of the total number $N_0$)
which propagate in the same channel where they were accepted up to the distance $z$ where 
they dechannel. 
The quantity $N_{\rm ch}(z)$ is the number of particles which are in the channeling mode 
irrespective of the channel which guides their motion at the distance $z$. 
With increasing $z$ the fraction $\xi_{\rm ch0}(z)$ decreases as the accepted electrons 
leave the entrance channel.
In the contrast, the fraction $\xi_{\rm ch}(z)$ 
can increase with $z$ when the electrons, including those not accepted at the entrance, 
can be captured in the channeling mode in the course of the  {rechanneling}. 
These dependencies for a 855 MeV electron channeling in straight and bent W(110) channels are 
presented in Fig. \ref{Figure.03} left.

A striking difference in the behaviour of the two fractions as functions of the 
penetration distance $z$ is mostly pronounced for the straight channel.
Away from the entrance point, the fraction $\xi_{\rm ch0}(z)$ (solid curve)
follows approximately the exponential decay law, 
$\xi_{\rm ch0}(z) \propto \exp\left(-z/L_{\rm p1}\right)$ (not shown in the figure).
At large distances, the fraction $\xi_{\mathrm ch}(z)$ (dashed curve) , which accounts for the 
rechanneling process, decreases much slower following the power law, 
$\xi_{\rm ch}(z)\propto z^{-1/2}$ \cite{Kostyuk-AK-AS-WG_855-Si}.
This dependence is shown in the figure with thick (red) dashed line.
As the bending curvature increases, $C\propto 1/R$, the rechanneling events become rarer, and
the difference between two fractions decreases.
For $C\gtrsim 0.1$ both curves virtually coincide.

Let us turn to the data on positron channeling presented in Table \ref{Table:ep-A-Lp12-Lch}
and  Fig. \ref{Figure.03} right.
In contrast to the electron case, the crystal length $L=75$ $\mu$m is much smaller than the 
positron dechanneling length in straight W(110) channel, $\Ld\approx 445$ $\mu$m, which can be 
obtained by means of the formula \cite{Dechan01,ChannelingBook2014,BiryukovChesnokovKotovBook}:
\begin{equation}
\Ld  
= \gamma {256 \over 9\pi^2} {\aTF \over r_0} {d\over \Lambda}.
\label{Results:eq.03}
\end{equation}
Here $r_0=2.8\times 10^{-13}$ cm is the electron classical radius, 
$\gamma=\E/mc^2$ is the Lorentz relativistic factor, 
$\aTF$ is the Thomas-Fermi atomic radius (equal to 0.112 \AA{} for a tungsten atom). 
The quantity $\Lambda$  stands for a so-called `Coulomb logarithm', which characterizes the 
ionization losses of an ultra-relativistic positron in an amorphous medium
(see e.g. \cite{Landau4}): $\Lambda = \ln\sqrt(2\gamma)^{1/2} mc^2/I - 23/24$
with $I$ being the mean atomic ionization potential ($I\approx 770$ eV for a tungsten atom).

To estimate the positron dechanneling length in a bent crystal, one can multiply
Eq. (\ref{Results:eq.03}) by the factor $(1-C)^2$ which exactly accounts for  
the relative change of the depth of the potential well in a bent channel
within the framework of the harmonic approximation for the continuous interplanar
potential \cite{BiryukovChesnokovKotovBook}.

For all values of $C$ indicated in Table \ref{Table:ep-A-Lp12-Lch}, the estimated dechanneling
length exceeds the crystal length $L$. 
Hence, the values of $L_{\rm p1}$ presented in the table can be
considered only as a lower bound of the positron dechanneling length.
For small bending parameters, $C\lesssim 0.1$, the penetration length is just below $L$,
thus signaling that most of the accepted projectiles traverse the whole crystal moving in the 
channeling mode (trajectory "p2" in Fig. \ref{Figure.01} is an exemplary one).
Finally, for all $C$, the total channeling length $\Lch$ is very close to the  
$\calA  L_{\rm p1}$.
Therefore, in contrast to the electron channeling, the rechanneling effect does not play 
any significant role even in the case of the straight crystal.
This statement is illustrated further by Fig. \ref{Figure.03}right where the 
fractions $\xi_{\rm ch0}(z)$ (solid curves) and $\xi_{\rm ch}(z)$ (symbols) 
are practically indistinguishable for $C>0$ and are very close for $C=0$.

\subsection{Radiation Spectra \label{Spectra}}

The simulated trajectories were also used to compute spectral distribution 
$\left\langle \d E/\hbar\d \om \right\rangle$ of the emitted 
radiation, Eq.  (\ref{Methodology:eq.03}).
For each trajectory, the distribution $\d^2 E_n/\d \om\, \d\Om$ was calculated accounting only
for the initial part of the trajectory with $z \leq 10$ $\mu$m.
Hence, the spectra discussed below refer to the $L=10$ $\mu$m thick crystals.
The integration over the emission angle $\theta$ was performed for 
two values of the radiation apertures:
$\theta_{0} = 0.24$ mrad and $\theta_{0} = 8$ mrad. 
The first value is close to that used in the experiments with the $855$ MeV electron 
beam at MAMI 
\cite{Backe_EtAl_2008,Backe_EtAl_2011,Backe_EtAl_PRL2014,Backe_EtAl_PRL_115_025504_2015}, 
and is much smaller than the natural emission angle for the beam energy, 
$\gamma^{-1} \approx 0.6$ mrad. 
Therefore, the corresponding spectra refer to a nearly forward emission. 
In contrast, the second aperture exceeds the $\gamma^{-1}$ 
by order of magnitude, thus providing the emission cone $\theta \leq \theta_{0}$ 
to collect almost all the radiation from the relativistic projectiles. 
The latter situation corresponds to the conditions at the experimental setup at 
SLAC~\cite{Wienands_EtAl_PRL_v114_074801_2015}. 

\begin{figure} [ht]
\centering
\includegraphics[width=7.5cm,clip]{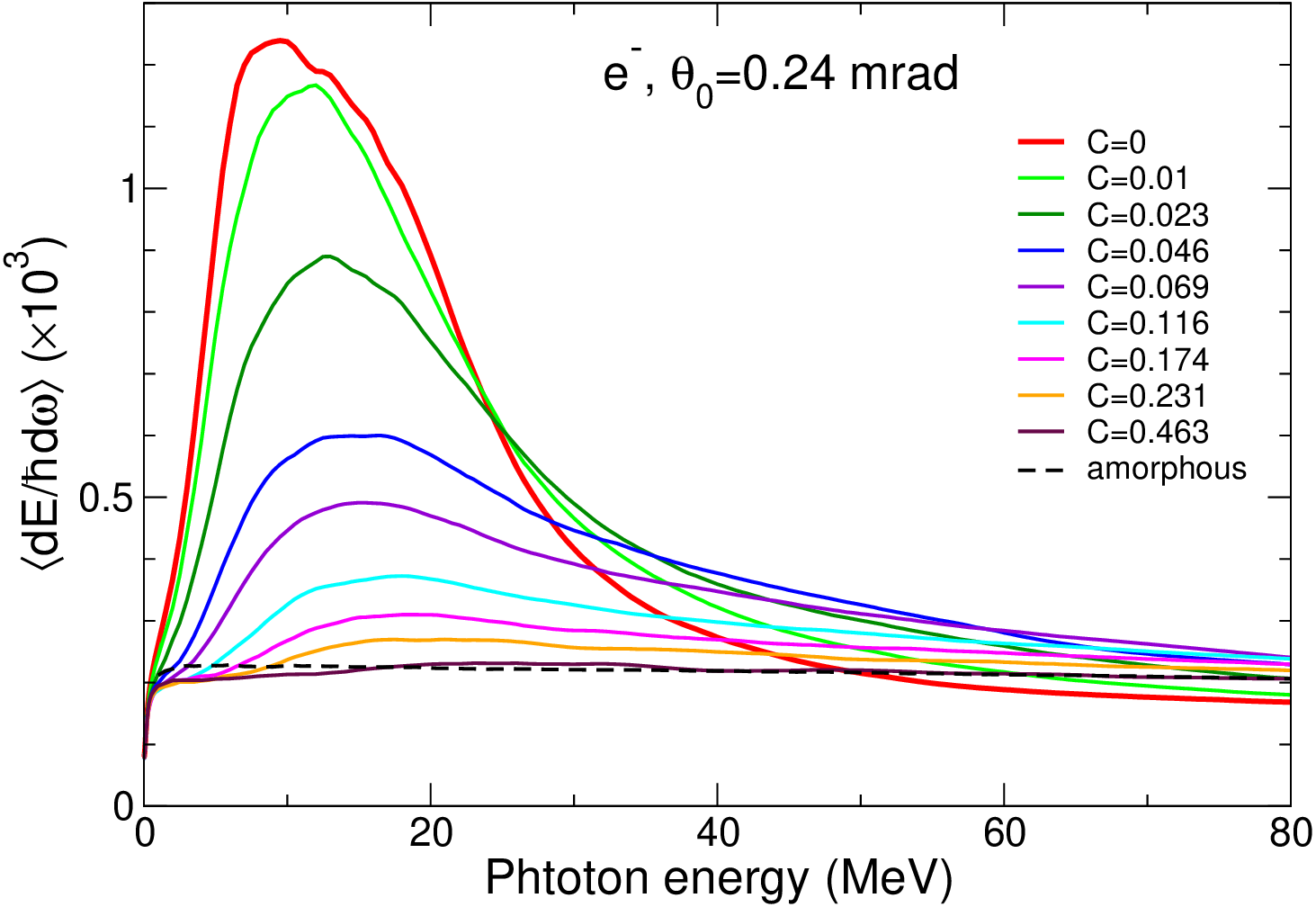}
\hspace*{0.5cm}
\includegraphics[width=7.5cm,clip]{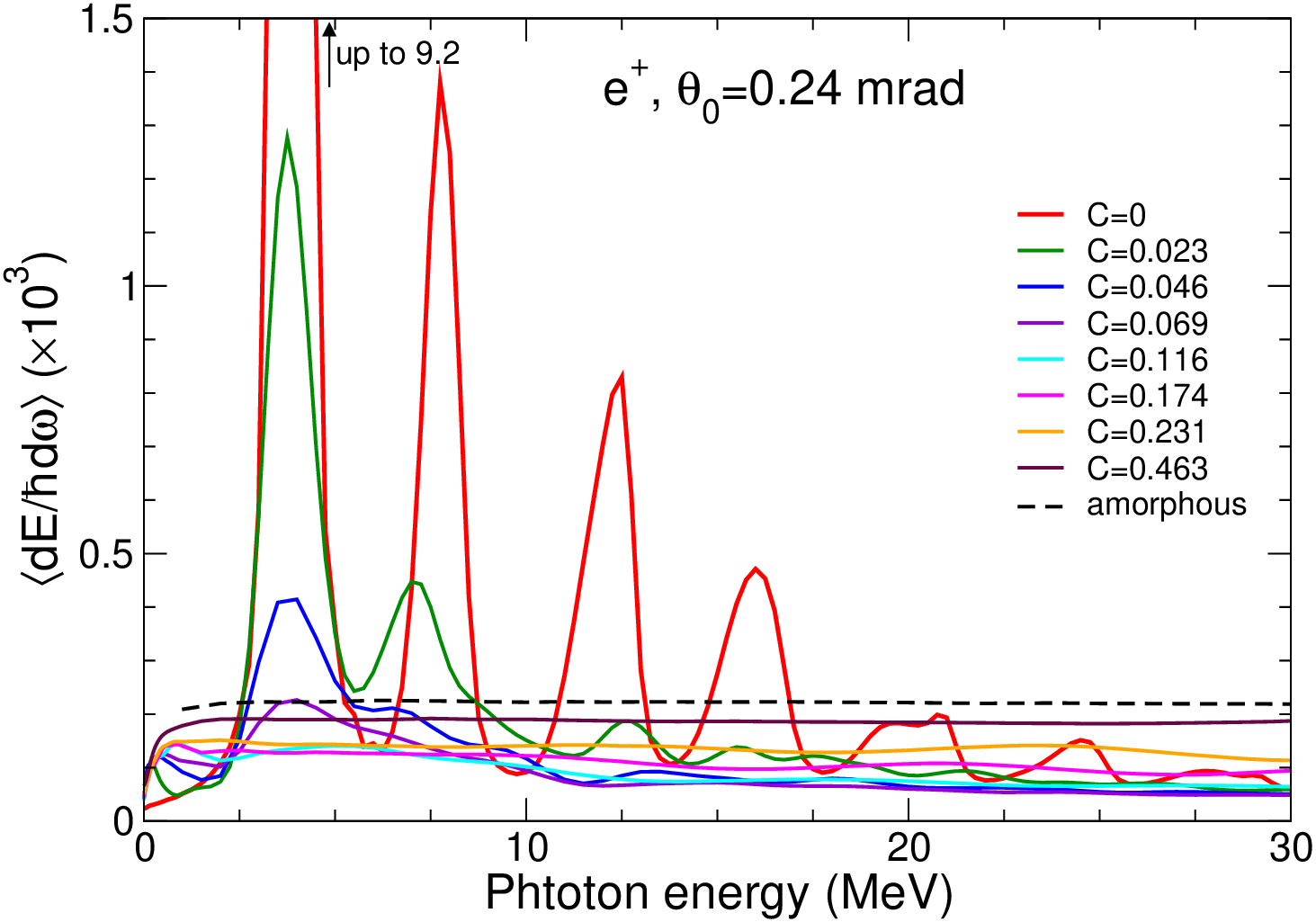}
\\
\vspace*{0.3cm}
\includegraphics[width=7.5cm,clip]{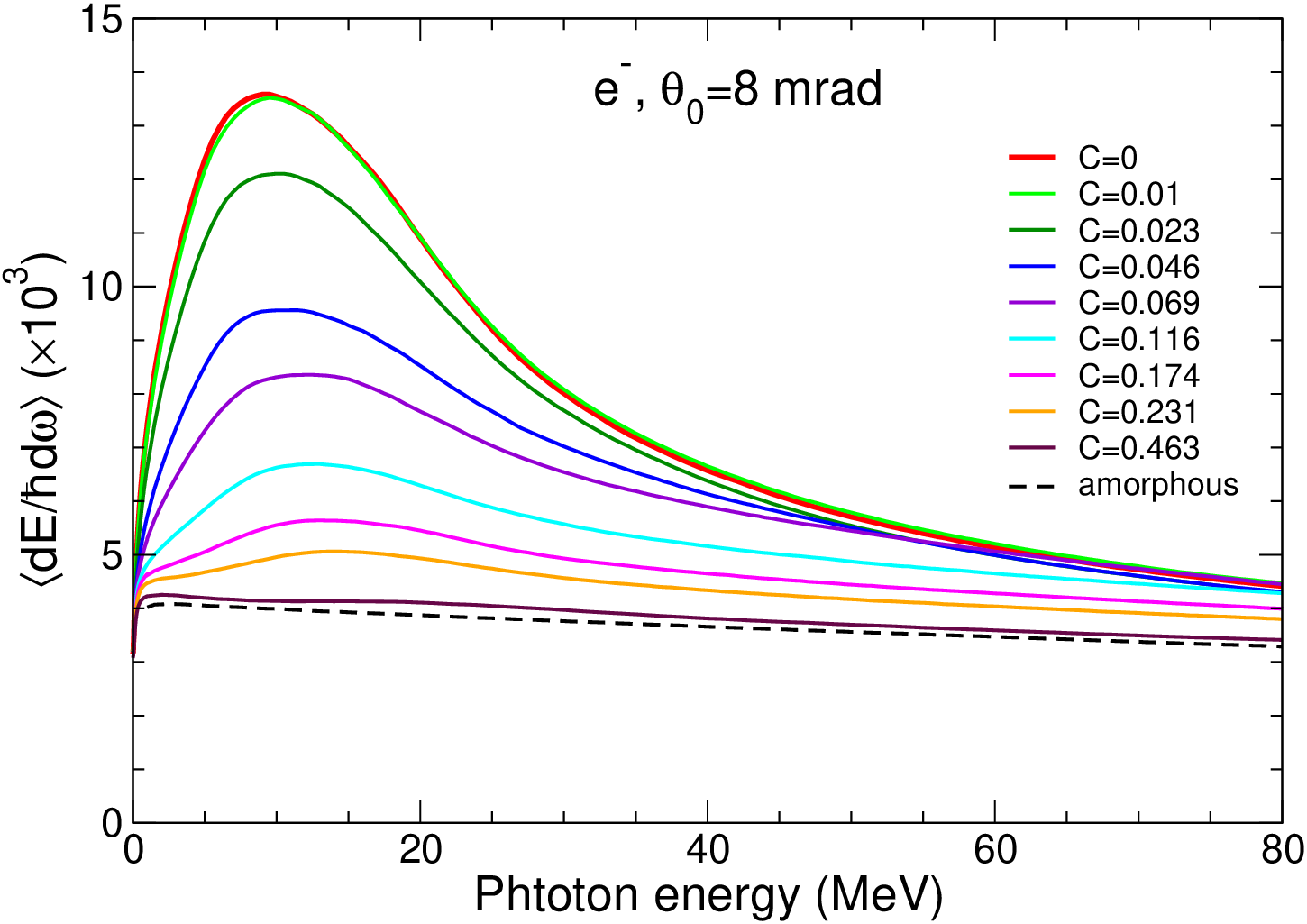}
\hspace*{0.5cm}
\includegraphics[width=7.5cm,clip]{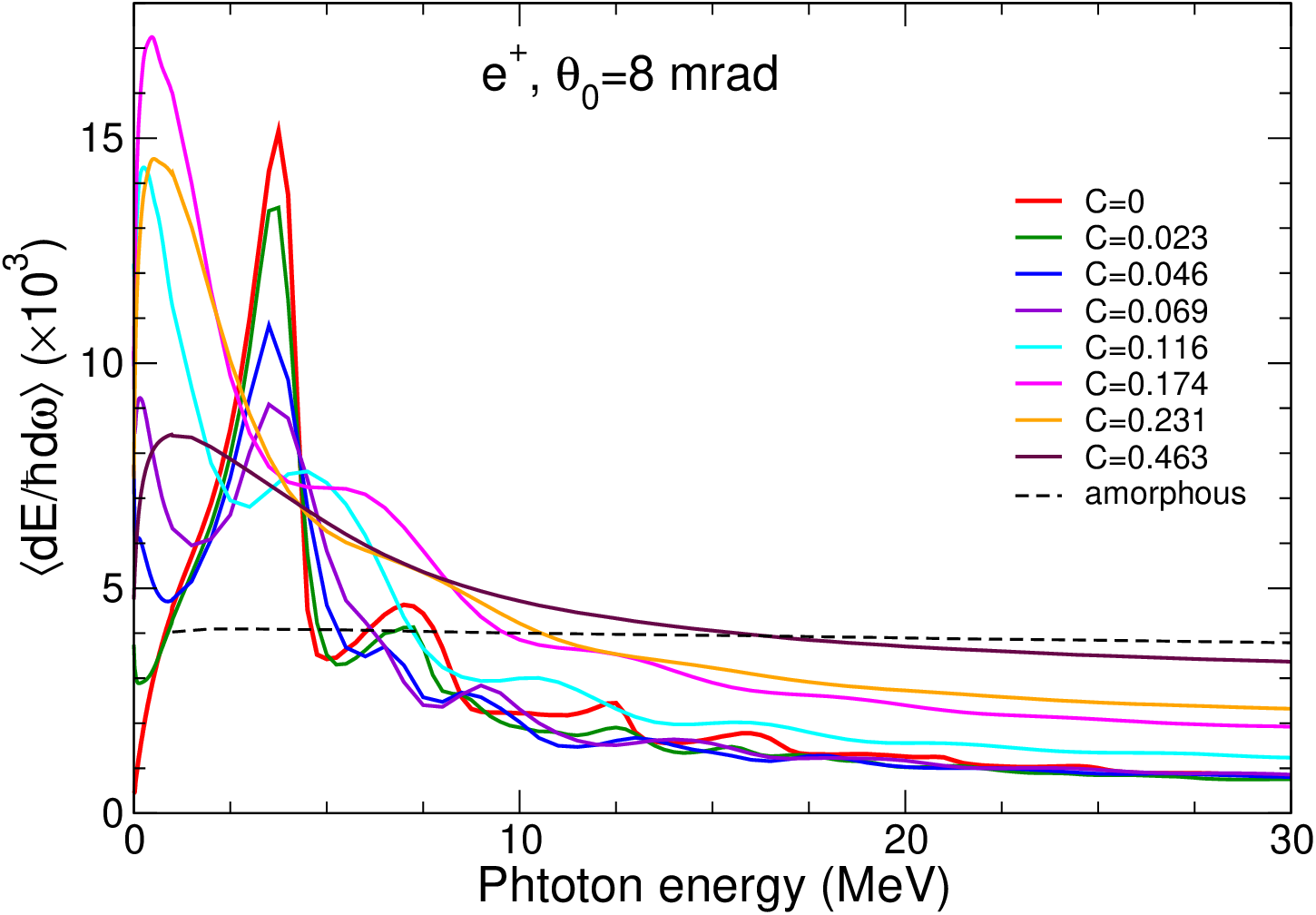}
\caption{
Spectral distribution of radiation emitted by 855 MeV electrons (left) 
and positrons (right) in straight (thick solid line, $C=0$) 
and bent (solid lines, $C>0$) W(110) crystals.
The dashed lines show the simulated spectra for amorphous tungsten. 
The upper and lower plots correspond to the aperture values 
$\theta_0 = 0.24$ and $8$ mrad, respectively. 
All spectra refer to the crystal thickness $L=10~\mu$m.}
\label{Figure.04}
\end{figure}

The plots in Fig. \ref{Figure.04} present the spectra emitted by electrons (left plots)
and positrons (right plots) within the apertures $\theta_{0}=0.24$ and 
$8$ mrad (top and bottom plots, respectively). 
In each plot, the thick solid (red) curve shows the spectrum in a straight 
W(110) channel (the bending parameter $C=0$), while other solid curves  
correspond to the bent channels ($C>0$, the corresponding values of the bending
radius one finds in Table \ref{Table:ep-A-Lp12-Lch}).
The dashed curves represent the spectra calculated from the simulated trajectories
in amorphous tungsten.

There are several features to be mentioned when comparing the emission
spectra calculated for two different apertures for each type of the projectile as
well as when comparing the electron and positron spectra.

To start with, we discuss the emission spectra from electrons.
In the case of a straight channel, 
the powerful maximum at $\hbar\om_{\max}\approx 10$ MeV
seen for both apertures (although more pronounced for $\theta_0 = 0.24$ mrad) is mainly 
due to the ChR.
In the maximum, the intensity emitted in the oriented W(110) crystal exceeds noticeably
the intensity $\langle \d E\rangle_{\rm am}$ of the incoherent BrS  
emitted in amorphous tungsten  (the dashed curves).
For bent crystals, the maximum of the intensity decreases with the increase of 
the bending parameter $C$.
Comparing the curves in the left-top and left-bottom panels of Fig. \ref{Figure.04}
one notices that, 
as $C$ increases, the ratio $\eta=\langle \d E\rangle_{C>0}/\langle \d E\rangle_{C=0}$ 
of the maximum intensities
in the bent and straight channels decreases faster for the smaller aperture.
For example, for $C=0.023$ the ratio equals to 0.7 for $\theta_0 = 0.24$ mrad but
$\eta=0.9$ for $\theta_0 = 8$ mrad.
This feature can be explain as follows.
In the vicinity of the maximum, the main contribution to the ChR intensity 
emitted within the cone $\theta\leq \theta_0$ comes from those channeling segments
for which the angle of inclination to the cone axis does not exceed $\theta_0$.
In a comparatively thin crystal, when $L<L_{\rm rech}$ \cite{footnote:01},
the rechanneling events are rare, so that the total channeling segment is 
determined by the initial penetration depth $L_{\rm p1}$.
In a straight crystal, the whole initial channeling segment is aligned with the 
emission cone so that the intensity is proportional to $L_{\rm p1}$.
In a bent channel, the depth $L_{\rm p1}$ should be compared to the scale $R\theta_0$.
The whole initial channeling segment can be considered
to be aligned with the cone axis if $L_{\rm p1} < R\theta_0$. 
In the opposite case, $L_{\rm p1} > R\theta_0$, the emission within the cone 
$\theta\leq \theta_0$  occurs effectively only from the part of the segment
\cite{BentSilicon_2014,Polozkov_VKI_Sushko_AK_AS_SPB_Diamond_2015}.
As a result, the maximum values of the intensities in the straight 
and bent crystals can be estimated as follows:
\begin{equation}
\left\langle \d E(\theta\leq \theta_0)\right\rangle_{C=0}
=
a \calA(0)  L_{\rm p1}(0)\,,
\qquad
\left\langle \d E(\theta\leq \theta_0)\right\rangle_{C>0}
=
a \calA(C) \min\left\{L_{\rm p1}(C),R\theta_0\right\}\,.
\label{Spectra:eq.01}
\end{equation}
Here the coefficient $a$ depends on the aperture, 
$\calA(C)$ and $L_{\rm p1}(C)$ stand for the acceptance and penetration depth
corresponding for a given bending parameter, see Table \ref{Table:ep-A-Lp12-Lch}.

For the larger aperture,
$\min\left\{L_{\rm p1}(C),R\theta_0\right\} = L_{\rm p1}(C)$ for all $C$ indicated.
Therefore, the ratio of the intensities is estimated as
$\eta_{l} \sim
\Bigl(\calA(C)/\calA(0)\Bigr)\,\Bigl(L_{\rm p1}(C)/L_{\rm p1}(0)\Bigr)$.
For the smaller aperture, $L_{\rm p1}(C) \leq R\theta_0$
only for $C\leq 0.015$ (which corresponds to $R\geq 1.3$ cm).
For larger bending parameters one evaluates 
$\eta_{s} \sim \Bigl(\calA(C)/\calA(0)\Bigr)\,\Bigl(R\theta_0/L_{\rm p1}(0)\Bigr) 
< \eta_{l}$.
 
To conclude the discussion of the electron spectra, let us comment on their behaviour
in the photon energy range well away from the maximum. 
It is seen that in most cases, the spectral distribution of radiation formed in 
the crystalline medium approaches from above that in amorphous tungsten.
The excess over $\langle \d E\rangle_{\rm am}$ is due to the emission of the 
coherent BrS by over-barrier particles which move 
along quasi-periodic trajectories, traversing the crystallographic planes
under the angle greater than Lindhard's critical angle 
(see Refs. \cite{Andersen1980,Ter-Mikaelian2001,Baier,AkhiezerShulga1982,%
BazylevZhevago,Sorensen1996,Uggerhoj_RPM2005} for the reviews on the coherent BrS).
The only exception from this scenario are the spectral dependencies
calculated for the smaller aperture in the straight ($C=0$) and nearly straight
($C=0.01$) crystals. 
In these cases, the amorphous background radiation is more intensive in the photon 
energy range  $\hbar\om \gtrsim 50$ MeV.
The explanation is as follows.
In a straight crystal, the coherent BrS is emitted by over-barrier projectiles
which move under the angles $\Theta>\Theta_{\rm L} = 0.56$ mrad with respect to the $z$-axis.
These projectiles mostly radiate within the cone $\gamma^{-1}$ along the instant
velocity so that only part of this radiation is emitted in the narrow cone 
$R\theta_0 =0.24$ mrad in the forward (with respect to the incident beam) direction.
In a bent crystal, a particle can become an over-barrier one, still moving along the 
initial $z$ direction, penetrating into the crystal at the distance $\gtrsim R\Theta_{\rm L}$.
Hence, for larger values of the bending parameter the intensity of the coherent
BrS emitted in the forward direction increases.

It is a well-established fact that channeling oscillations of electrons have a 
strong anharmonic character which is a direct consequence of a strong deviation of 
the electron interplanar potential from a harmonic shape (see, e.g., \cite{BakEtal1985}).
The period of oscillations varies with the amplitude, as it is illustrated by two 
simulated electron trajectories presented in Fig. \ref{Figure.01}.
As a result, the spectral distribution of ChR  exhibit a rather broad maximum.
  
In contrast, the channeling trajectories of positrons demonstrate nearly harmonic oscillations 
between the neighboring planes. 
This is also in accordance with a well-known result established within the framework of the 
continuum model of channeling \cite{Gemmel}. 
Indeed, for a positively charged projectile the interplanar potential can be 
approximated by parabola in most part of the channel thus leading to close resemblance 
between the channeling motion of positrons and the undulator motion.
As a result, for each value of the emission angle $\theta$ the spectral distribution
of ChR in a straight crystal  
reveals a set of narrow and equally spaced peaks (harmonics).
The harmonic frequencies, $\om_n$, of ChR of positrons can be estimated from
\begin{eqnarray}
\om_{n}
=
{2\gamma^2\, \Om_{\rm ch} \over 1 + \gamma^2 \theta^2 + K_{\rm ch}^2/2}
\, n ,
\quad
n=1,2,3,\dots .
\label{Spectra:eq.02}
\end{eqnarray}
Here, $\Om_{\rm ch}$ and $K_{\rm ch}$ are the frequency 
and the undulator parameter of the channeling oscillations.
The maximum value of the latter can be estimated as 
$2\pi\gamma(d/2)/\lambda_{\rm ch}$ with $d/2$ and $\lambda_{\rm ch}$ being, respectively,
the maximum possible amplitude and the period of the oscillations.
Within the framework of harmonic approximation for the interplanar potential, one derives 
$\Om_{\rm ch}=2d/c\Theta_{\rm L}$ and 
$K_{\rm ch} \leq \gamma\Theta_{\rm L}$. 
For a 855 MeV positron channeling in straight W(110) crystal this estimate produces 
$K_{\rm ch} \lesssim 0.9$. 

It is known from the theory of undulator radiation (see, e.g., \cite{AlferovBashmakovCherenkov1989}) 
that for $K\sim 1$ the emission spectrum contains few harmonics the intensities of which
rapidly decrease with the harmonic number $n$.
This feature is explicitly seen in the spectral distributions calculated from the 
simulated trajectories of positrons propagating in {\rm straight} W(110), see the 
thick solid curves in the right plots in Fig. \ref{Figure.04}.
The well-defined peaks (more pronounced for the smaller aperture) correspond to the 
harmonics of the ChR.
For both apertures, the most powerful first peak is located at $\approx 4$ MeV.
This value corresponds to the energy of the first (or, fundamental) harmonic
emitted in the forward direction which one obtains from Eq. (\ref{Spectra:eq.02})
by setting $n=1$, $\theta=0$ and using the aforementioned estimates for 
$\Om_{\rm ch}$ and $K_{\rm ch}$.
The intensities of the emission into higher harmonics (the peaks with $n$ up to 5 are
seen located at $\hbar\om_n\approx 4n$ MeV) rapidly decrease with $n$.
This harmonic-like structure of the spectral distribution of ChR of positrons 
is clearly distinguishable from smooth curve which characterizes the electron 
spectrum of ChR.

The data presented in Table \ref{Table:ep-A-Lp12-Lch} for the straight channel 
allows one to compare the maximum intensities
of ChR of positrons and electrons in the straight $L=10$ $\mu$m thick crystal.
For electrons, the peak intensity can be estimated directly from the first equation 
in (\ref{Spectra:eq.01}).
For positrons, the penetration depth $L_{\rm p1}(0)$ exceeds greatly the crystal thickness,
therefore, to estimate the peak intensity one substitutes $L_{\rm p1}(0)$ with $L$.
As a result, one obtains the following estimate for the ratio of the intensities:  
$\left\langle \d E(\theta\leq \theta_0)\right\rangle_{C=0}^{\rm pos}/
\left\langle \d E(\theta\leq \theta_0)\right\rangle_{C=0}^{\rm el} 
= \calA_{\rm pos}(0)  L /(\calA(0)  L_{\rm p1}(0))_{\rm el} \approx 5$
which correlates with the value which can be calculated by comparing the peak intensities
in the top plots in Fig. \ref{Figure.04}.  

As the bending curvature increases, the spectral distributions of radiation emitted
by positrons become modified following two different scenarios.

The first one manifests itself as the decrease in the peak intensities with the increase 
of $C$. 
This feature is much more pronounced for the smaller aperture, see      
top-right plot in Fig. \ref{Figure.04}.
For example, the intensity of the first harmonic peak for $C=0.026$ is six times less than
that in the straight crystal.
This decrease rate is much larger than in the case of electron channeling where the 
corresponding drop is on the level of 30 per cent.
Similar effect is seen for the higher harmonics as well as for larger values of $C$.

The explanation is as follows. 
The data on the penetration depth $L_{\rm p1}$ for positrons, presented in 
Table \ref{Table:ep-A-Lp12-Lch}, indicate that $L_{\rm p1}>L=10$ $\mu$m for 
all considered values of $C$.
Hence, on average, all accepted particles propagate through the crystal moving in the
channeling mode.
As mentioned, in a straight crystal, the peak intensity can be estimated from 
the first equation in (\ref{Spectra:eq.01}) where one substitutes 
$L_{\rm p1}(0)$ with the crystal length $L$.
However, even for the $C$ value as small as 0.023, the emission into the nearly forward cone 
$\theta_0=0.24$ mrad occurs from the much shorter initial part of the channeling
trajectory of the length $R\theta_0\approx 2.1$ $\mu$m.
Hence, the estimate of the intensities ratio reads:  
$\left\langle \d E(\theta\leq \theta_0)\right\rangle_{C=0.026} /
\left\langle \d E(\theta\leq \theta_0)\right\rangle_{C=0}
= \calA(0.23)  R\theta_0 / \calA(0)  L \approx 1/5$, which correlates with the data from
top-right plot in Fig. \ref{Figure.04}.

For the larger aperture, $\theta_0=8$ mrad, the scale $R\theta_0$ exceeds the crystal length $L$
for $C$ up to $0.16$.
Hence for $C\leq 0.16$ the ratio of the intensities is mainly determined by the 
ratio of the acceptances, $\calA(C)/\calA(0)$ which drops off at a much lower rate.
This tendency one sees when comparing the peak intensities of the spectral distributions 
shown in bottom-right plot in Fig. \ref{Figure.04}.   
 
\begin{figure} [ht]
\centering
\includegraphics[width=10cm,clip]{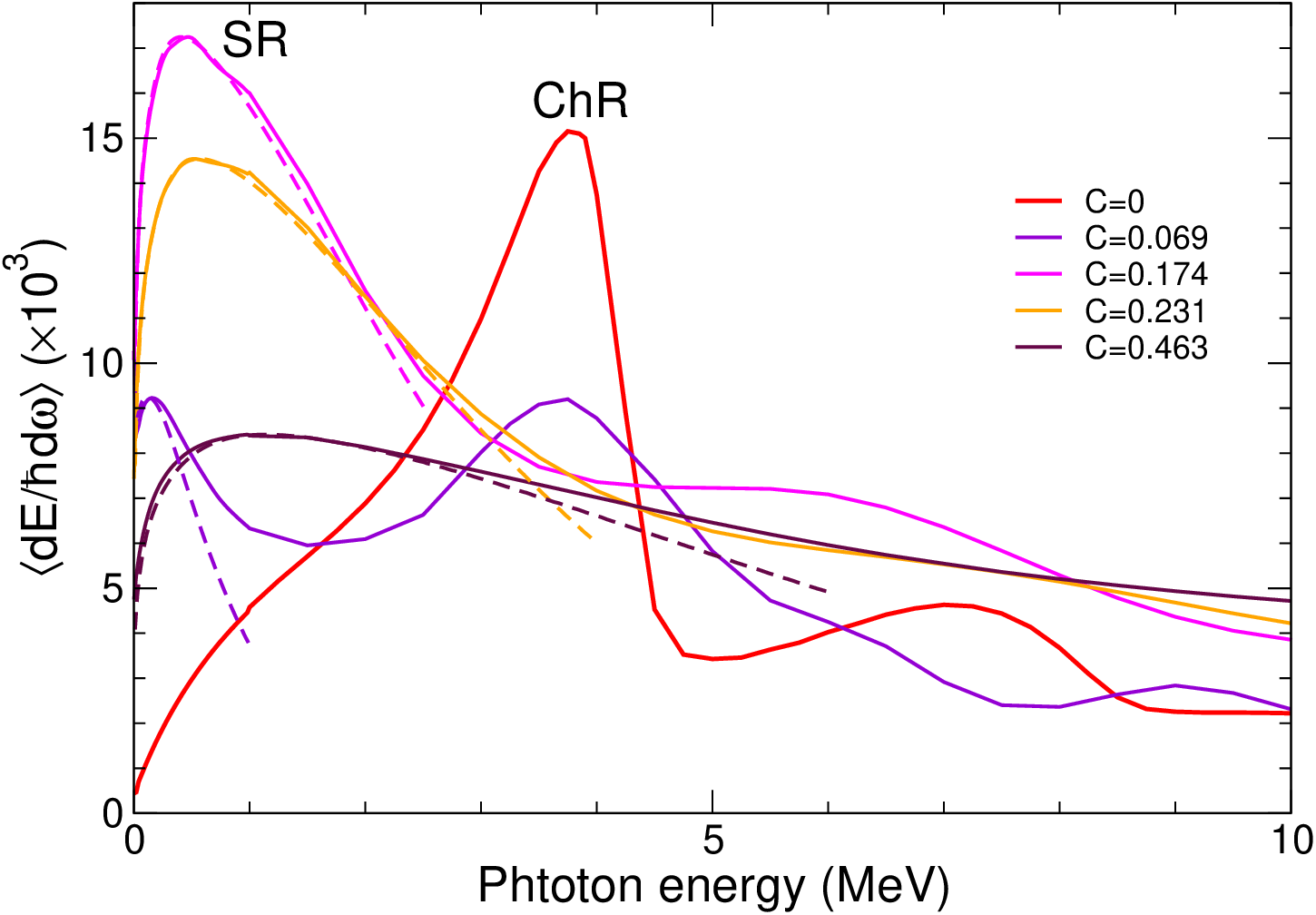}
\caption{
Same as in bottom-right plot in Fig. \ref{Figure.04} but for smaller photon energy range and 
for several representative values of the bending parameter $C$. 
The peaks at $\hbar\om \leq 1$ MeV are due to the synchrotron radiation (SR, dashed curves). 
See also explanation in the text.
}
\label{Figure.05}
\end{figure}

The second scenario of the spectral intensity modification with the increase 
of the bending parameter $C$ is due to the contribution of the synchrotron radiation (SR)
to the emission spectrum.
This feature is most pronounced for the larger aperture, 
see bottom-right plot in Fig. \ref{Figure.04}. 

Positrons, which channel along the planes in a bent crystal experience two
types of motion: the channeling oscillations and the translation along the centerline 
of the bent channel.
The latter motion gives rise to the synchrotron-type radiation, i.e. the one, which
is emitted by a charge moving along a circle (or, its part) (see, e.g., \cite{Jackson}).
Therefore, the total spectrum contains the features of both ChR and SR.
It was shown in Ref.\,\cite{Bashmakov1981}, that in the case of planar channeling
the crystal bending noticeably affects the spectrum only if the condition
$\Theta_{\rm L}(C) \gamma \leq 1$ is met ($\Theta_{\rm L}(C)$ is the critical angle in the bent
channel).
In this case, the total spectrum preserves the features of ChR if
$L_{\rm c} \gg \lambda$ (here, $L_{\rm c} = R/\gamma$ 
is the formation (coherence) length of the
radiation emitted from an arc of the circle of the radius $R$, and $\lambda$ is the 
radiation wavelength), but becomes
of the quasi-synchrotron type in the opposite limit \cite{Bashmakov1981,Taratin_Review_1998}.
The peculiarities which appear in spectral and spectra-angular distributions
of the emitted radiation due to the interference of the two mechanisms of radiation
were analyzed in Refs.\,\cite{SolovyovSchaeferGreiner1996,ArutyunovEtAl_NP_1991,%
TaratinVorobiev1988,TaratinVorobiev1989} 
by analytical and numerical means.
In Ref. \cite{BentSilicon_2014} the  results of numerical simulations of the
emission spectra by 855 MeV electrons were reported for straight and uniformly bent 
Si(110) crystal. 
The influence of the detector aperture on the form of the spectral distribution of the 
emitted radiation in bent S(110) was explored in \cite{Sub_GeV_2013}.
The results of simulation of planar channeling as well as the calculated 
spectral distribution of the emitted radiation were reported in 
\cite{Sushko_EtAl_NIMB_v355_p39_2015} for 3\dots20 GeV electrons and positrons in 
bent Si(111).

The curves in bottom-right plot in Fig. \ref{Figure.04} show that  
for $C>0$ the SR manifests itself as an additional structure in the 
low-energy part of the spectrum  ($\hbar\om \lesssim 1$ MeV).  
The intensity of the SR peak increases with the bending parameter up to $C\sim L/\theta_0$, and
then it decreases due to the reason discussed above: the larger curvature is, the smaller is
the part of the trajectory which contributes to the radiation cone $\theta_0$.
To clearer visualize the relationship of the additional structure and the SR, 
we present  Fig. \ref{Figure.05} where the spectra calculated from the simulated trajectories 
for several representative values of $C$ (solid curves) are plotted in a narrower 
range of photon energies.
Also plotted are the spectral distributions of SR  (dashed curves) emitted by 855 MeV 
positrons moving along the circular arc of the length $L=10$ $\mu$m and of the   
bending radius $R$ corresponding to the indicated value of $C$
(see Table \ref{Table:ep-A-Lp12-Lch}).
These dependencies were scaled to match the simulated distributions in the maxima 
of the SR.

\section{Conclusions}

By means of the channeling module \cite{ChanModuleMBN_2013} of the \MBNExplorer
package \cite{MBN_ExplorerPaper,MBN_ExplorerSite}, we have performed the 
simulations of classical trajectories of 855 MeV electrons and positrons in $L=75$ $\mu$m thick 
oriented straight and bent single tungsten crystals.
The energy considered is sufficiently large to disregard quantum effects when describing
the interaction of the projectile with the constituent atoms.
Therefore, the relativistic classical mechanics framework chosen in the paper is  
fully adequate.
The trajectories were analyzed to quantify the channeling process by calculating various
parameters (acceptance, penetration length, total channeling length) as functions of the 
curvature.
The obtained results were compared to the data available. 

The simulated trajectories were used as the input data for numerical analysis of the intensity
of the emitted radiation in 10 $\mu$m thick crystals. 
In the case of straight crystals the channeling radiation appears atop the incoherent 
bremsstrahlung background. 
In a bent channel, the spectrum is enriched by the synchrotron radiation due to the circular 
motion of the projectile along the bent centerline. 
In all cases, the spectral  distribution of radiation was calculated for two extreme values 
of the detector aperture $\theta_0$ as compared to the natural angle $\gamma^{-1}$ of the 
emission cone by an ultra-relativistic projectile.
The smaller aperture, $\theta_0 = 0.24$ mrad, being much less than $\gamma^{-1}$
(equal to 0.60 mrad for a 855 electron and/or positron), 
characterizes the radiation spectrum emitted in the (nearly) forward direction. 
This aperture is routinely used in the channeling experiments with electrons at 
the Mainz Microtron facility 
\cite{Backe_EtAl_2008,Backe_EtAl_PRL_115_025504_2015}.
The larger aperture, $\theta_0 = 8$ mrad, accounts for nearly all emitted radiation 
in the case of straight and moderately bent crystals. 

The obtained and presented results are of interest in connection with the ongoing experiments 
with crystalline undulators at MAMI 
\cite{Backe_EtAl_2011,BackeLauth_Dyson2016,Backe_EtAl_PRL2014}
carried out with electron beams as well as  
with possible experiments by means of the positron beam \cite{Backe_EtAl_2011a}.
In these experiments the silicon-based crystalline undulators were exposed 
(or discussed to be exposed) to the beam.
However, of current interest is the search for other crystalline structures 
capable to effectively steer ultra-relativistic light projectiles along 
periodically bent crystallographic planes \cite{ChannelingBook2014}.
Therefore, the results presented in the current paper 
can serve as benchmark data for further theoretical and experimental efforts
in studying channeling and radiation formed in tungsten-based crystalline undulators.

\section*{Acknowledgments}

This work is supported by the National Natural Science Foundation of China under 
Grant Nos. 11025524 and 11161130520, 
the National Basic Research Program of China under Grant No. 2010CB832903,
 and 
by the European Commission (the PEARL Project within the H2020-MSCA-RISE-2015 call, 
GA 690991). 
Constructive suggestions by Alexey Verkhovtsev are gratefully acknowledged.

\section*{References}


\begin{thebibliography}{99}
\bibitem{footnote:01}
        We remind, that the spectral dependencies discussed here refer to 
        $L=10$ $\mu$m which is smaller that the rechanneling length $L_{\rm rech}=20$ $\mu$m 
        estimated above for the straight crystal.
\end{thebibliography}
\end{document}